\documentclass[
  aps, prb,
  reprint,
  amsmath,
  floatfix,
  ]{revtex4-2}

\usepackage[T1]{fontenc}
\usepackage{libertinus}
\usepackage{libertinust1math}
\usepackage[cal=stixtwoplain, bb=pazo]{mathalpha}

\usepackage{microtype}
\usepackage{mathtools}
\usepackage{bm}

\usepackage{placeins}

\usepackage{xcolor}
\usepackage{graphicx}
\graphicspath{{./img/}}

\usepackage{enumitem}
\usepackage{booktabs}

\let\xtimes\times
\let\times\cdot
\usepackage{csquotes}
\usepackage[
  input-comparators={<=>\approx\ge\geq\gg\le\leq\ll\sim\lesssim},
  range-phrase={\,\text{--}\,},
  product-symbol=\xtimes,
  ]{siunitx}
\usepackage[version=4]{mhchem}

\usepackage{hyperref}
\usepackage[all]{hypcap}
\usepackage[ocgcolorlinks]{ocgx2}

\makeatletter
  \patchcmd\frontmatter@setup{\normalfont}{\normalfont\sffamily}{}{}
  \patchcmd\frontmatter@title@format{\bfseries}{\sffamily\bfseries}{}{}
  \patchcmd\frontmatter@affiliationfont{\it}{\slshape}
  \patchcmd\frontmatter@title@format{\large}{\Large}{}{}
  \patchcmd\section{\bfseries}{\sffamily\bfseries}{}{}
  \patchcmd\subsection{\bfseries}{\sffamily\bfseries}{}{}
  \patchcmd\subsubsection{\itshape}{\sffamily}{}{}
  \patchcmd\@makecaption{\rmfamily}{\sffamily}{}{}
\makeatother

\makeatletter
  \NewDocumentCommand\titlecaption{ m m }{%
    \caption[#1]{\textbf{#1.}\hskip 1.5\fontdimen2\font plus 1em minus 1.5\fontdimen4\font\relax#2}%
  }
  \pretocmd\@make@capt@title{\begingroup\bfseries}{}{}
  \patchcmd\@make@capt@title{\@caption@fignum@sep}{\@caption@fignum@sep\endgroup}{}{}
  \def\@caption@fignum@sep{: }

\makeatother

\makeatletter
  \newcommand*\@withperiod[1]{#1.}
  \patchcmd\paragraph{\normalfont\normalsize\itshape}{\normalfont\normalsize\itshape\@withperiod}{}{}
\makeatother

\makeatletter
  \def\bibsection{%
    \expandafter\section\expandafter*\expandafter{\refname}%
    \@nobreaktrue
    \raggedright
  }%
\makeatother

\makeatletter
  \AddToHook{env/table/begin}{%
    \skip@\abovecaptionskip
    \abovecaptionskip\belowcaptionskip
    \belowcaptionskip\skip@
  }
\makeatother

\usepackage{figutil}

\definecolor{hyperintcolor}{rgb:HTML}{416b16}
\definecolor{hyperextcolor}{rgb:HTML}{2263a3}
\hypersetup{
  colorlinks,
  linkcolor=hyperintcolor,
  citecolor=hyperintcolor,
  filecolor=hyperextcolor,
  urlcolor=hyperextcolor,
}

\RenewDocumentCommand\doi{ m }{%
  \href{\doibase #1}{\nolinkurl{#1}}%
}

\def\e{\mathrm{e}}
\def\i{\mathrm{i}}

\let\vec\bm

\makeatletter
  \def\mathopsym#1{\mathop{\kern\z@#1}}
  \DeclareRobustCommand\Olandau{\mathopsym{\mathcal O}}
  \NewDocumentCommand\abs{ m }{%
    \lvert#1\rvert
  }
  \NewDocumentCommand\ket{ m }{%
    \lvert#1\rangle
  }
  \NewDocumentCommand\bra{ m }{%
    \langle#1\rvert
  }
  \NewDocumentCommand\estate{ m }{%
    \ensuremath{\mathrm{#1}}%
  }
  \def\@difsym{{\mathrm d}}
  \def\@delsym{\partial}
  \NewDocumentCommand\dif{s s o m}{%
    \IfBooleanF{#2}{\IfBooleanTF{#1}{\unskip}{\mathop{}}\!}%
    \@difsym%
    \IfValueT{#3}{^{#3}}%
    #4%
    \IfBooleanT{#1}{\mathop{}\!}%
  }
  \NewDocumentCommand\del{o m}{%
    \@delsym_{\mkern-2.5mu\relax#2}%
    \IfValueT{#1}{^{#1}}%
  }
  \NewDocumentCommand\evalwith{O{\big} m}{%
    \mathop{}\!#1|_{#2}%
  }
\makeatother

\DeclareMathOperator\erf{erf}

\newcommand*\Gammasc{\Gamma_{\!\mathrm{sc}}}

\DeclareSIUnit\sqr{\ensuremath{\square}}

\NewDocumentCommand\sublabel{ m }{(#1)}
\NewDocumentCommand\subref{ m m }{\ref{#1}\,\sublabel{#2}}

\NewDocumentCommand\software{}{\textit}
\NewDocumentCommand\lisaplus{}{LISA\textsuperscript{\!\bfseries +}}

\AddToHook{env/table/begin}{\advance\abovecaptionskip 1.8pt\relax}
\AtBeginDocument{%
  \abovedisplayshortskip=5pt plus 2pt minus 5pt\relax
}

\begin{document}

\title{A superconducting on-chip microwave cavity\\for tunable hybrid systems with optically trapped Rydberg atoms}

\def\pitaffil{%
  \affiliation{Physikalisches Institut, Center for Quantum Science~(CQ) and \lisaplus, Universität Tübingen, 72076~Tübingen, Germany}
}
\author{Benedikt~Wilde}
\email{benedikt.wilde@uni-tuebingen.de}
\pitaffil
\author{Manuel~Kaiser}
\pitaffil
\author{Malte~Reinschmidt}
\pitaffil
\author{Andreas~Günther}
\pitaffil
\author{Dieter~Koelle}
\pitaffil
\author{Jószef~Fortágh}
\pitaffil
\author{Reinhold~Kleiner}
\pitaffil
\author{Daniel~Bothner}
\email{daniel.bothner@uni-tuebingen.de}
\pitaffil

\begin{abstract}
  Hybrid quantum systems are highly promising platforms for addressing important challenges of quantum information science and quantum sensing.
  Their implementation, however, is technologically non-trivial, since each component typically has unique experimental requirements.
  Here, we work towards a hybrid system consisting of a superconducting on-chip microwave circuit in a dilution refrigerator and optically trapped ultra-cold atoms.
  Specifically, we focus on the design optimization of a suitable superconducting chip and on the corresponding challenges and limitations.
  We unfold detailed microwave-cavity engineering strategies for maximized and tunable coupling rates to atomic Rydberg-Rydberg transitions in \ce{^87Rb}~atoms while respecting the boundary conditions due to the presence of a laser beam near the surface of the chip.
  Finally, we present an experimental implementation of the superconducting microwave chip and discuss the cavity characteristics as a function of temperature and applied dc~voltage.
  Our results illuminate the required consideration aspects for a flexible, tunable superconductor-atom hybrid system, and lay the groundwork for realizing this exciting platform in a dilution refrigerator with vacuum Rabi frequencies approaching the strong-coupling regime.
\end{abstract}

\makeatletter
  \appto\titleblock@produce{\addvspace{-2\p@}}
\makeatother

\maketitle

\section{Introduction}
Coupling superconducting microwave circuits to other physical platforms such as spins, phonons, quantum dots or atoms has the potential to enable a large variety of groundbreaking hybrid quantum technologies~\cite{xiang2013, kurizki2015, clerk2020}.
Hybrid architectures are particularly promising for compensating intrinsic weaknesses of specific quantum systems by the strengths of the other components.
A prominent example are hybrid systems with superconducting qubits~\cite{clerk2020, clarke2008, blais2021} -- the circuits provide fast and reliable quantum processors, but suffer from short coherence times and from a lack of technologies to reliably transmit microwave-photonic quantum states over long distances.
Both challenges could be solved by coupling the superconducting circuits to e.g.~mechanical oscillators or to atoms, which in principle can both serve as quantum memories~\cite{rabl2006, verdu2009, zhao2009, li2016, wallucks2020, liu2023} and microwave-to-optical transducers~\cite{andrews2014, han2018, petrosyan2019,  kumar2023}.
However, combining different platforms in a single device and with sufficiently large coupling rates to facilitate coherent state transfer between them is typically a highly non-trivial experimental task.
In this work, we focus on the specific aspect of interfacing neutral atoms with a superconducting on-chip microwave resonator; in particular on the design of a suitable and optimized superconducting chip for the coupling to ultracold Rydberg atoms.

Different possibilities to couple superconducting cavities to atoms have been explored in the past.
The most famous is probably the method of coupling individual flying Rydberg atoms to a three-dimensional superconducting microwave cavity, an ultra-low-loss photon-box, which allowed for groundbreaking and unprecedented control over photonic quantum states~\cite{haroche2013}.
Rydberg atoms are ideal for achieving large coupling rates to microwave photons since they possess large electric transition dipole moments~\cite{soerensen2004, petrosyan2008, petrosyan2009, hogan2012}.
Similar experimental settings with propagating atom beams are lately being investigated using on-chip microwave cavities, although with much higher cavity losses and lower coupling rates~\cite{thiele2015, morgan2020, walker2020, walker2022}.
A second approach is interfacing on-chip microwave circuits with atomic ensembles trapped nearby~\cite{boehi2009, bernon2013, hattermann2017, kaiser2022}, which has the advantage of atomic (quasi-)stationarity and a potentially large collective enhancement of otherwise small interaction rates~\cite{dicke1954, rabl2006, verdu2009}.
The atom-trapping can be achieved either magnetically on (microwave) atom-chips~\cite{boehi2009, fortagh2007} or with optical tweezers~\cite{grimm2000, gard2017, covey2019}, i.e.~a strongly focused laser beam which is red-detuned from a suitable atomic resonance.
Magnetic trapping with superconductors is accompanied by complications in microwave-chip design~\cite{bernon2013, bothner2013} due to the necessity for on-chip trap-current leads and Meissner distortion of external magnetic fields~\cite{cano2008}.
Optical trapping close to the chip surface does not come with these constraints and has been discussed in various proposals~\cite{beck2016, gard2017}, but has yet to be demonstrated.
It does, however, imply various technological challenges related to combining lasers with a \unit{\milli\kelvin}~system and superconductors, that need to be considered with care.
Independent of the trapping method, one of the most critical aspects of a potentially useful superconductor-atom hybrid system is the single-photon coupling rate -- the vacuum Rabi frequency -- which in all existing on-chip implementations has been orders of magnitude below the strong-coupling regime~\cite{hattermann2017, morgan2020, kaiser2022}.

\begin{figure*}
  \includegraphics{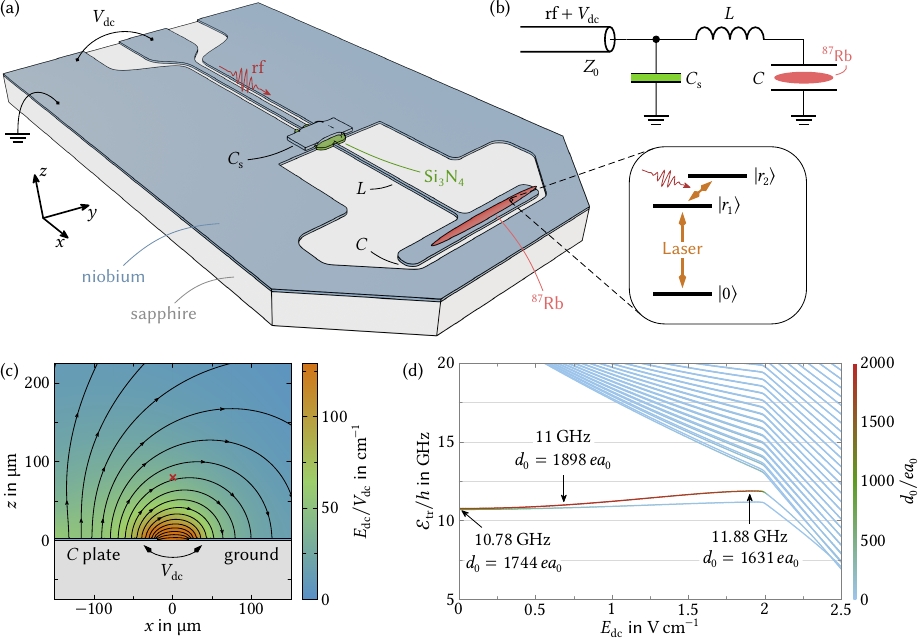}
  \titlecaption{A superconducting on-chip microwave cavity with dc voltage access for coupling to optically trapped Rydberg atoms}{%
    \sublabel{a}~Schematic of the envisioned hybrid system.
    On top of a sapphire substrate (light gray), a superconducting microwave resonator is patterned from a thin layer of niobium (Nb, blue).
    The resonant LC~circuit comprises an inductance~$L$ and a capacitance~$C$.
    At one of its ends, the circuit is coupled to a coplanar waveguide~(CPW) feedline by a trilayer shunt capacitor $C_{\mathrm s}$, and at its other end to an ensemble of ultra-cold rubidium~\ce{^87Rb} atoms.
    The atoms (red cigar) are levitating $\qty{\sim 80}{\micro\meter}$ above the capacitor gap of the circuit, confined in an optical dipole trap.
    The center of the capacitor gap at the substrate surface is at $\vec r = (x, y, z) = (0, 0, 0)$.
    \sublabel{b}~Equivalent circuit of the microwave resonator with a resonance frequency $\omega_0 = 1 / \sqrt{L C_{\mathrm{tot}}}$, where $C_{\mathrm{tot}}^{-1} = C^{-1} + C_{\mathrm s}^{-1}$.
    The CPW feedline has a characteristic impedance~$Z_0 \approx \qty{50}{\ohm}$.
    Due to the shunt-coupling to the feedline, it is possible to apply a dc~bias voltage $V_{\mathrm{dc}}$ to the circuit capacitance~$C$ in which the atomic ensemble is placed.
    Using a laser (not shown), the atoms can be excited from their electronic ground-state~$\ket 0$ to the Rydberg manifold~$\ket{r_i}$, where Rydberg-Rydberg transitions with large electric dipole moments and frequencies in the \unit{\giga\hertz}~regime can couple to the superconducting resonator.
    Here, we focus on a transition with a transition frequency $\bigl( \mathcal E_2 - \mathcal E_1 \bigr) / h \approx \qty{11}{\giga\hertz}$, where $h$ is Planck's constant and $\mathcal E_i = \mathcal E(\ket{r_i})$.
    \sublabel{c}~Cross-sectional schematic in the vicinity of the position of the atomic ensemble with $y = 0$ and the numerically simulated ratio of dc~electric field strength~$E_{\mathrm{dc}}$ to applied voltage~$V_{\mathrm{dc}}$ versus $x$-$z$-position; electric field lines are indicated.
    Values close to the gap edges that exceed the scale of the colorbar are drawn in the orange of the maximum.
    The resulting field at the location of the atomic ensemble (marked~\plotref{plt:cloudpos}\,) is $E_{\mathrm{dc}} = V_{\mathrm{dc}} \times \qty{37}{\per\centi\meter}$.
    \sublabel{d}~Differential Stark map showing transition frequencies $\mathcal E_{\mathrm{tr}} / h = \bigl( \mathcal E - \mathcal E_1 \bigr) / h$ within the rubidium Rydberg manifold in the vicinity of the zero-field transition $\ket{\estate{59P_{3/2}}, m_j = \frac 32} \to \ket{\estate{58D_{5/2}}, m_j = \frac 32}$.
    The corresponding transition dipole moments~$d_0$ in units of $e a_0$ ($e$~is the elementary charge, $a_0$~is the Bohr radius) are indicated by color.
    The red line marks the high-$d_0$ target transition~$\ket{r_1} \to \ket{r_2}$.
    At $E_{\mathrm{dc}} = \qty{0.68}{\volt\per\centi\meter}$, which corresponds to $V_{\mathrm{dc}} = \qty{18.7}{\milli\volt}$, it has a transition frequency of \qty{11}{\giga\hertz} and $d_0 = \num{1898} \, e a_0$.
  }
  \label{fig:design-concept}
\end{figure*}

Here, we report the development of an on-chip superconducting microwave cavity which has been optimized for the electric dipole-coupling to Rydberg transitions in optically trapped ultra-cold atomic ensembles.
The chosen resonance frequency is around~$\qty{11}{\giga\hertz}$, which is directly compatible with state-of-the art superconducting qubits.
In addition to the microwave field, our resonator design enables the application of a dc electric field for tuning the Rydberg transitions into and out of resonance as well as compensating stray electric fields from the chip surface, e.g.~due to adsorbates~\cite{hattermann2012, chan2014, kaiser2022, ocola2024}.
We discuss the geometric boundary conditions that allow the chips to be approached by a high-intensity focused laser beam and devise the layout of the circuit based on the expected profile of the resulting atom-trap.
As a next step, we demonstrate how to optimize the geometric details of the superconducting resonator for achieving large vacuum Rabi frequencies and reveal that even with conservative parameter estimates the strong-coupling regime is within reach.
Finally, we present an implementation of the optimized microwave chip and show experimental results of its characterization in a cold-atom dilution refrigerator.
We discuss in particular the dependence of cavity properties on temperature and applied dc~voltage.
Our device and results pave the way for a new generation of tunable superconductor-atom hybrid systems in a \unit{\milli\kelvin}~environment and in the strong-coupling regime.

\section{Results}
\subsection{Preliminary considerations}
We begin by visualizing our experimental concepts and by defining the boundary conditions and the chosen parameter space for the envisioned superconductor-atom hybrid system, cf.~Fig.~\ref{fig:design-concept}.
Our final objective is to implement the system in a dilution refrigerator, since only then the microwave circuits will be in their quantum ground state.
The specific cold-atom dilution refrigerator under consideration has been presented in Ref.~\cite{jessen2014}, and throughout the present manuscript we will respect the restrictions given by this cryogenic system.
These include the size of the windows for optical access to the \unit{\milli\kelvin}~plate and the presence of a magnetic conveyor belt to transfer the atoms from the position of the magneto-optical trap~(MOT) at~\qty{6}{\kelvin} to the mixing chamber plate, where the superconducting chip is mounted.

The target hybrid system provides a variety of exciting perspectives, such as enabling novel on-chip quantum experiments with Rydberg atoms~\cite{patton2013, pritchard2014, sarkany2015, sarkany2018}, advances in quantum frequency-conversion~\cite{han2018, petrosyan2019, covey2019, lauk2020, kumar2023, petrosyan2024}, hybrid quantum gates and the realization of atomic quantum memories~\cite{pritchard2014, petrosyan2009}, atom number counting~\cite{stammeier2017, garcia2019}, cavity cooling~\cite{sarkany2018}, and electric field sensing~\cite{sedlacek2012, thiele2015, walker2022}.
Since essentially all of these experiments will considerably benefit from a high microwave-cavity--Rydberg interaction strength, a high gate speed and gate fidelity as well as from protocol simplicity, we focus on a system with a 2D~microwave circuit coupled to a pair of Rydberg states with an easily preparable ground state and a large transition dipole moment.
Particularly interesting would be the single-atom--single-photon strong coupling regime, which enables e.g.~quantum-coherent single-excitation state transfer between the microwave cavity and a single Rydberg atom, and could enable the hybridization of superconducting qubits and Rydberg atoms using the cavity as a quantum bus in the next step~\cite{petrosyan2009, yu2017, semiao2025}.
Compared to their 3D~counterparts, 2D~microwave circuits can provide orders of magnitude larger zero-point microwave fields due to their compact mode volume, but at the expense of a much lower quality factor (assuming the same material), which explains the necessity for the use of a superconductor.
Speed and simplicity are furthermore important since the system already has a high experimental complexity by default and it contains various low-frequency noise and decoherence sources such as atom motion, atom losses, voltage noise, laser phase noise and position vibrations of the chip~/ the laser beam, induced e.g.~by the dilution refrigerator pulse tube.
Further advantages of a chip-based implementation are easy access to the microwave cavity for the dipole trap, low photon absorption probability on the \unit{\milli\kelvin}~components, small Meissner-distortion of the magnetic trapping fields during atom preparation, and compactness of the complete microwave-part, which (at least in our system) has to fit into the center of a split-coil at the end of a magnetic conveyor belt (cf.~Appendix~\ref{sec:app_setup} and Ref.~\cite{jessen2014}).
Finally, our chosen approach will leave sufficient space and access to potentially add the components for a single-atom detector~\cite{kaiser2022} in a later experimental stage.

We start by considering the most important requirements of our envisioned hybrid system.
First and foremost, the superconducting chip and in particular its packaging need to allow optical access at the location of the atoms.
Secondly, the atoms should be trapped as close to the chip surface as possible to maximize the interaction with the superconducting cavity.
The atoms should also be confined to a small region to minimize interaction inhomogeneities across the ensemble~\cite{hattermann2017, kaiser2022}.
Direct interaction of trapping-laser photons with the superconducting chip, however, should be avoided, since this would lead to Cooper pair breaking or even to heating of the entire mixing chamber to which the chip is anchored.
We conclude that a system like the one shown in Fig.~\subref{fig:design-concept}{a} provides the best balance in addressing these initial requirements:
An optical dipole trap (laser not shown) is used to levitate an atomic cloud $\qty{\sim 80}{\micro\meter}$ above the chip surface~\cite{laser-chip-distance-note} and close to the edge of a tip-tapered superconducting chip.
The result is similar to the approach in Ref.~\cite{beck2016}, albeit completely planar and devoid of normal metal elements.
Our design also takes the elongated ensemble distribution in a fridge-compatible dipole trap into account.
The details of the exact shape and size of the chip, the shape of the atomic cloud and the chip-atom distance are discussed in Secs.~\ref{sec:dipole-trap} and~\ref{sec:parameter-optimization}.

The ellipsoidal shape of the atomic ensemble in combination with the objective of maximized vacuum Rabi frequencies dictate some of the requirements for the layout details of the superconducting resonator.
Firstly, all the ensemble atoms should be able to participate in the interaction with equal coupling strength; this reduces inhomogeneous broadening and increases the collective coupling~\cite{kubo2012, julsgaard2013}.
It is therefore essential that the resonator and microwave fields are translationally invariant along the longitudinal axis of the ensemble.
Secondly, since we target an electric dipole interaction, all the electric microwave fields should ideally be concentrated in the region of the atoms.
Hence, the optimal layout is a lumped-element LC~circuit with the capacitance~$C$ directly below the position of the atoms and a minimal amount of stray capacitances.
Based on these factors, we devised the T-shaped microwave cavity shown in Fig.~\subref{fig:design-concept}{a}:
A single straight inductance~$L$ is connected to a coplanar capacitance~$C$ to ground; to minimize the stray capacitance, the ground plane has a large distance to the circuit elements except in the interaction region with the atoms.
At this point it also becomes clear why coplanar waveguide cavities -- as often used in theoretical proposals and in other experimental approaches~\cite{petrosyan2008, verdu2009, bernon2013, bothner2013, pritchard2014, beck2016, hattermann2017, morgan2020, walker2020, walker2022, kaiser2022, petrosyan2024} -- are not ideal for maximized coupling and field homogeneity.
By their very nature as distributed element circuits, they come with significantly distributed (i.e.~stray) capacitances and spatially varying microwave electric fields along the transmission line.

As third design aspect, it would be highly beneficial to have the possibility to apply a dc electric field~$E_{\mathrm{dc}}$ to the atomic ensemble.
This can not only be used to tune the Rydberg transition into and out of resonance with the cavity, but also to (partially) compensate for static parasitic adsorbate fields.
To allow for this option, we couple the microwave cavity to its coplanar waveguide~(CPW) feedline by means of a parallel-plate shunt capacitor~$C_{\mathrm s}$ to ground~\cite{bosman2015}.
In this coupling scheme the CPW center conductor is not interrupted and remains galvanically connected to $L$ and~$C$, and a dc electric field above the chip can just be realized by applying a corresponding dc~voltage~$V_{\mathrm{dc}}$ to the feedline center conductor, cf.~Fig.~\subref{fig:design-concept}{c}.
If necessary, the direction of the dc~field can be adjusted from $x$- to $z$-direction by moving the dipole trap and the atoms along~$x$.
Generating the dc~electric field via the cavity itself has the elegant side-effect of obtaining (nearly) identical dc and microwave field configurations, which could be used to pre-select a slice of resonant atoms in the cloud which then all couple with identical Rabi frequencies to the cavity photons~\cite{kaiser2022}.

Finally, we need to design the cavity for a specific resonance frequency that matches a chosen Rydberg-Rydberg transition of \ce{^87Rb}~atoms.
For frequency-compatibility with the most common superconducting quantum circuits and common microwave equipment we decided to operate in the frequency range \qtyrange{\sim 10}{12}{\giga\hertz}.
In addition to a matching transition frequency, a suitable Rydberg-Rydberg transition needs to have a large electric dipole transition matrix element and be sufficiently distant from other such transitions.
We choose the transition $\ket{\estate{59P_{3/2}}, m_j = \frac 32} \to \ket{\estate{58D_{5/2}}, m_j = \frac 32}$ with a transition frequency $\mathcal E_{\mathrm{tr}} / h \approx \qty{11}{\giga\hertz}$ and a transition dipole moment of $d_0 \approx 1800 \, e a_0$, the exact values being dependent on the dc electric field.
A differential Stark map around the corresponding transition is shown in Fig.~\subref{fig:design-concept}{d}, illustrating that it possesses the required properties.
It has been numerically calculated using the methods described in Refs.~\cite{grimmel2015, kaiser2022}, cf.~Appendix~\ref{sec:app_starkmap}.

\subsection{Of lasers and atoms}\label{sec:dipole-trap}
\begin{figure}
  \includegraphics{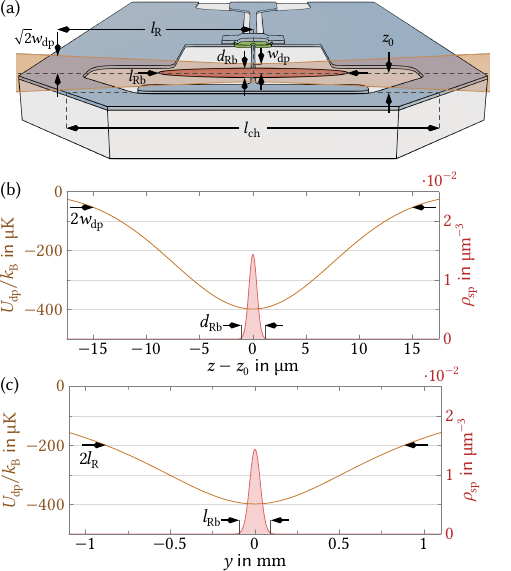}
  \titlecaption{Position and trapping characteristics of the atoms near the superconducting microwave chip}{%
    \sublabel{a}~Schematic of the chip tip in front view, viewing direction is~$-x$ with a small elevation angle, cf.~Fig.~\ref{fig:design-concept}.
    Above the chip surface, in a distance~$z_0$, the ensemble is trapped by a focused laser beam (dipole trap).
    The beam waist ($1/\e^2$~radius) in the focus plane is denoted as~$w_{\mathrm{dp}}$ and the Rayleigh length of the beam focus as~$l_{\mathrm R}$.
    For a chip width underneath the laser axis~$l_{\mathrm{ch}} = \qty{1.6}{\milli\meter}$ we choose $z_0 = \qty{80}{\micro\meter}$ and $w_{\mathrm{dp}} = \qty{15}{\micro\meter}$, which corresponds to $l_{\mathrm R} = \qty{0.88}{\milli\meter}$.
    In \sublabel{b} and~\sublabel{c} we show the resulting trapping potential $U_{\mathrm{dp}} / k_{\mathrm B}$ for ground-state rubidium atoms in $z$- and $y$-direction, respectively, using a laser with wavelength $\lambda_{\mathrm{dp}} = \qty{800}{\nano\meter}$ and power $P_{\mathrm{dp}} = \qty{50}{\milli\watt}$.
    More details on the calculation and the parameters used can be found in the main text.
    For a realistic assumption of an atomic ensemble temperature $T_{\ce{Rb}} \sim \qty{1}{\micro\kelvin}$, we get the Gaussian single-atom density profiles~$\rho_{\mathrm{sp}}$ (red lines and shaded areas) overlaid to the potentials in \sublabel{b} and~\sublabel{c}.
    The width of the atomic cloud is $d_{\ce{Rb}} = \qty{2.3}{\micro\meter}$ in $z$-direction (as well as in $x$-direction) and $l_{\ce{Rb}} = \qty{0.19}{\milli\meter}$ in $y$-direction (we define the size of the distribution as six standard deviations).
  }
  \label{fig:dipole-trap}
\end{figure}

Since it defines the further layout of the chip and the resonator, we calculate the shapes of the focused laser beam and of the atomic ensemble trapped by it before considering aspects of the chip design on a more quantitative level.
Several critical length scales are competing with each other here and a satisfying compromise between the beam waist at the focus, the angle of divergence and the distance from the chip surface has to be found, cf.~Fig.~\subref{fig:dipole-trap}{a}.
As laser wavelength for the dipole trap we choose $\lambda_{\mathrm{dp}} \approx \qty{800}{\nano\meter}$, which is red-detuned from the rubidium $\estate{D_1}$~($\ket{\estate{5S_{1/2}}} \to \ket{\estate{5P_{1/2}}}$) and $\estate{D_2}$~($\ket{\estate{5S_{1/2}}} \to \ket{\estate{5P_{3/2}}}$) transitions at \qty{795}{\nano\meter} and~\qty{780}{\nano\meter}, respectively.
For a given laser intensity $I_{\mathrm{dp}}(\vec r)$, which is a function of position $\vec r$, the trapping potential can be calculated by~\cite{grimm2000}
\begin{equation}\label{eq:dipole_potential}
  \begin{aligned}[b]
    U_{\mathrm{dp}}(\vec r)
    &= \biggl( -\frac{\pi c_0^2}{2 \omega_1^3} \, \Bigl( \frac{\Gamma_1}{\omega_1 - \omega_{\mathrm{dp}}} + \frac{\Gamma_1}{\omega_1 + \omega_{\mathrm{dp}}} \Bigr) \\
    &\phantom{ = \biggl( } -\frac{\pi c_0^2}{\omega_2^3} \, \Bigl( \frac{\Gamma_2}{\omega_2 - \omega_{\mathrm{dp}}} + \frac{\Gamma_2}{\omega_2 + \omega_{\mathrm{dp}}} \Bigr) \biggr) \, I_{\mathrm{dp}}(\vec r) ,
  \end{aligned}
\end{equation}
where $c_0$ is the vacuum speed of light, $\omega_{\mathrm{dp}}$ is the laser frequency, $\omega_1 = 2 \pi \times \qty{377}{\tera\hertz}$ and $\omega_2 = 2 \pi \times \qty{384}{\tera\hertz}$ are the frequencies and $\Gamma_1 = 2 \pi \times \qty{5.75}{\mega\hertz}$ and $\Gamma_2 = 2 \pi \times \qty{6.07}{\mega\hertz}$ are the linewidths of the atomic transitions $\estate{D_1}$ and $\estate{D_2}$, respectively~\cite{steck2024rubidium87d}.

The intensity distribution of a focused Gaussian laser beam propagating in $y$-direction and focused in the $y = 0$ plane is given by
\begin{equation}\label{eq:dipole_intensity}
  I_{\mathrm{dp}}(\vec r) = \frac{2 P_{\mathrm{dp}}}{\pi w^2(y)} \, \exp\Bigl( -\frac{2 r^2}{w^2(y)} \Bigr)
\end{equation}
with $r = \sqrt{x^2 + \tilde z^2}$ the radial distance from the focus center in the $x$-$z$-plane.
Here $\tilde z \coloneq z - z_0$ is the $z$-distance from that focus center, which is positioned at a distance~$z_0$ from the chip surface.
Thus, the overall shape of the trapping potential is that of a Gaussian in $x$- and $z$-direction and that of a Lorentzian in $y$-direction.
The $1/\e^2$~radius of the beam~$w(y)$ is given by
\begin{equation}
  w(y) = w_{\mathrm{dp}} \sqrt{1 + \frac{y^2}{l_{\mathrm R}^2}}
\end{equation}
with the beam waist in the focus plane~$w_{\mathrm{dp}}$.
The Rayleigh length~$l_{\mathrm R}$ appearing here is not independent of the beam waist, but given by
\begin{equation}\label{eq:Rayleigh_length}
  l_{\mathrm R} = \frac{\pi w_{\mathrm{dp}}^2}{\lambda_{\mathrm{dp}}} .
\end{equation}
It defines the length in propagation direction on which the beam radius increases by a factor of~$\sqrt 2$ from $w_{\mathrm{dp}}$ at~$y = 0$.
Considering the dimensions of the cryogenic setup and the windows for optical access, we can comfortably work with $w_{\mathrm{dp}} = \qty{15}{\micro\meter}$, which corresponds to $l_{\mathrm R} = \qty{0.88}{\milli\meter}$.
Together with the width of the superconducting chip underneath the laser beam axis, these values provide a limit for how close to the surface the atoms can be trapped while ensuring that only minor amounts of laser radiation reach the chip.

The next question for designing cavity and chip is how large the atom cloud in that trap will be.
The two main parameters defining the size are the ensemble temperature~$T_{\ce{Rb}}$, which we expect to be on the order of \qty{\sim 1}{\micro\kelvin}, and the laser power~$P_{\mathrm{dp}}$.
With $P_{\mathrm{dp}} = \qty{50}{\milli\watt}$ we find a maximum trap depth of $T_{\mathrm{dp}} = \abs{U_{\mathrm{dp}}} / k_{\mathrm B} \approx \qty{400}{\micro\kelvin}$, cf.~Fig.~\ref{fig:dipole-trap}, and for the much colder atoms the trap shape constitutes in good approximation a harmonic oscillator potential in all three dimensions.
The atomic distribution in a 3D harmonic potential is a 3D Gaussian~\cite{eschner2003}, and the single-particle density profile is described by
\begin{equation}\label{eq:atom_density}
  \rho_{\mathrm{sp}}(r, y) = \frac{1}{\sqrt{2 \pi}^3 \sigma_r^2 \sigma_y} \, \exp\Bigl( -\frac{r^2}{2 \sigma_r^2} - \frac{y^2}{2 \sigma_y^2} \Bigr) .
\end{equation}
The radial and longitudinal variances $\sigma_r^2$ and $\sigma_y^2$ are
\begin{equation}
  \sigma_q^2 = \frac{k_{\mathrm B} T_{\ce{Rb}}}{m_{\ce{Rb}} \omega_q^2} , \quad q \in \{r, y\}
\end{equation}
with $m_{\ce{Rb}}$ the mass of a rubidium atom, $k_{\mathrm B}$ Boltzmann's constant, and $\omega_r$ ($\omega_y$) the radial (longitudinal) harmonic oscillation frequency, cf.~Appendix~\ref{sec:app_oscfreqs}.
From the shape of our beam we finally obtain a cloud extension of $d_{\ce{Rb}} \coloneq 6 \sigma_r = \qty{2.3}{\micro\meter}$ in radial direction and $l_{\ce{Rb}} \coloneq 6 \sigma_y = \qty{0.19}{\milli\meter}$ in longitudinal direction, cf.~also Fig.~\ref{fig:dipole-trap}.
To be on the safe side and have sufficient margin even for elevated cloud temperatures of up to $T_{\ce{Rb}} \sim \qty{5}{\micro\kelvin}$ we set the length of the capacitor plate along the laser axis to be fixed at $l = \qty{1}{\milli\meter}$.

Two remaining, mutually dependent values that we need to determine are the width~$l_{\mathrm{ch}}$ of the chip underneath the laser axis and the distance~$z_0$ of the atoms from the chip surface.
It is intuitively clear from Fig.~\ref{fig:dipole-trap} that a wider chip requires a larger atom-chip distance, at least if we want to avoid the tail of the Gaussian laser beam to shine on the superconductor.
As a satisfying compromise including some margin for error we find a chip width of~$l_{\mathrm{ch}} = \qty{1.6}{\milli\meter}$ and a chip-atom distance of $z_0 = \qty{80}{\micro\meter}$.
These values guarantee completely negligible direct interaction between the laser photons and the superconductor, cf.~Appendix~\ref{sec:app_laser_on_chip}, while still providing enough space for capacitor, gap and ground plane as discussed above.
Since on the other hand a chip with a constant width of~$\qty{1.6}{\milli\meter}$ is not convenient for mounting and wire bonding, and does not provide much space for minimizing stray capacitances, we have opted as an ideal solution for a tapered-tip design.
Regarding the atom-chip distance, we will experimentally explore the possibility to reduce it further in the future, although there might be additional aspects to consider such as parasitic stray fields that can get very strong close to the surface~\cite{hattermann2012, chan2014, kaiser2022, ocola2024}.

We note at this point that due to the large parameter space to consider and various experimental options (laser wavelength, laser intensity, beam profile, ensemble temperature) there might exist even more ideal configurations to be discovered.
Our considerations, however, describe an instructive way for how to engineer a satisfying compromise based on specific boundary conditions and experimental and technological choices.
That being said and the atom position being fixed, we are ready to optimize the details of the superconducting microwave resonator for maximized interaction rates in the next section.

\subsection{Maximizing the coupling rate}\label{sec:parameter-optimization}

\begin{figure*}
  \includegraphics{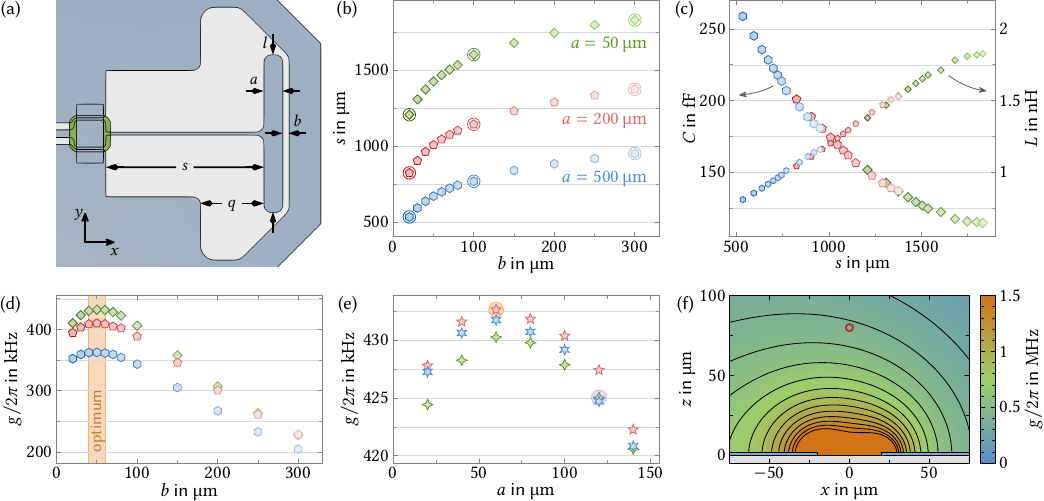}
  \vskip -4pt\relax
  \titlecaption{Optimizing the geometry parameters of the superconducting cavity for maximized coupling strength}{%
    \sublabel{a}~Schematic of the chip in top view with definition of all relevant resonator parameters used in the optimization.
    The length of the inductance wire is~$s$, the width of the capacitor plate is~$a$, and the gap width between plate and ground-plane at the far end is~$b$.
    The length of the capacitor plate $l = \qty{1}{\milli\meter}$ is a fixed parameter chosen to accommodate atomic cloud lengths of up to $\qty{\sim 0.8}{\milli\meter}$ (cf.~Fig.~\ref{fig:dipole-trap}).
    Also, the general shape of the ground plane cutout including the parameter $q = \qty{400}{\micro\meter}$ is kept unchanged during the optimization process.
    If we change either $a$ or~$b$ to optimize the coupling rate to the atoms, the capacitance~$C$ changes accordingly and we need to adjust~$s$ to keep $\omega_0 = 2 \pi \cdot \qty{11}{\giga\hertz}$.
    In~\sublabel{b} we present the resulting~$s$ for $a \in \{ \plotref{plt:l50} \, \qty{50}{\micro\meter}, \plotref{plt:l200} \, \qty{200}{\micro\meter}, \plotref{plt:l500} \, \qty{500}{\micro\meter} \}$ and various~$b$.
    The value of~$b$ is additionally indicated by a color intensity gradient, which is also applied to the corresponding data in panels \sublabel{c} and~\sublabel{d}.
    The values for~$s$ are obtained by an iterative simulation procedure using \software{Sonnet} for the circled values and by interpolation for the other values (see main text).
    \sublabel{c}~shows the corresponding $C$ and $L = 1 / \omega_0^2 C$ (the inductance values having smaller marks for better differentiation) for each parameter set $(a, b, s)$ from~\sublabel{b}.
    The values for~$C$ are obtained from \software{Comsol Multiphysics} simulations which simultaneously provide the electric field strength at the position of the atomic ensemble center $(x, y, z) = (0, 0, 80) \, \unit{\micro\meter}$ for a given voltage across the capacitor.
    This field strength in turn can be used to calculate the coupling rate $g = E_{\mathrm{zpf}} d_0 / \hbar$, where $d_0$ is the transition dipole matrix element and $E_{\mathrm{zpf}} = \abs{\vec E_{\mathrm{zpf}}}$ is the root-mean-square field amplitude of the resonator quantum ground state.
    \sublabel{d},~\sublabel{e}~Coupling rate~$g$ at the center of the atom cloud for various~$a$ and~$b$ (with $d_0 = 1898e a_0$).
    In~\sublabel{d} we use the same values for~$a$ as before, in~\sublabel{e} we focus on $b \in \{ \plotref{plt:b40} \, \qty{40}{\micro\meter}, \plotref{plt:b50} \, \qty{50}{\micro\meter}, \plotref{plt:b60} \, \qty{60}{\micro\meter} \}$, since that is the region of strongest coupling as indicated by the orange interval in~\sublabel{d}.
    The maximum value $g_{\mathrm{max}} = 2 \pi \cdot \qty{433}{\kilo\hertz}$ is obtained for $a = \qty{60}{\micro\meter}$ and $b = \qty{50}{\micro\meter}$ (orange highlighted data point).
    The device presented in this paper was fabricated with $a = \qty{120}{\micro\meter}$ and $b = \qty{40}{\micro\meter}$ (violet highlighted data point), corresponding to $g = 2 \pi \cdot \qty{425}{\kilo\hertz} = 0.98 g_{\mathrm{max}}$.
    \sublabel{f}~$g$~as a function of position $(x, y = 0, z)$ near the capacitor gap at the end of a resonator with $a = \qty{120}{\micro\meter}$, $b = \qty{40}{\micro\meter}$ and $s = \qty{1.1}{\milli\meter}$.
    Contour lines mark curves of equal coupling strength at the values indicated by ticks in the colorbar.
    Values close to the gap edges that exceed the scale of the colorbar are drawn in the orange of the maximum.
    The position of the atomic ensemble is marked by a red circle; the circle diameter is chosen as~$2 d_{\ce{Rb}}$ (cf.~Fig.~\ref{fig:dipole-trap}) to enhance its visibility.
  }
  \label{fig:parameter-optimization}
\end{figure*}

The optimization of the resonator layout for large interaction rates can be condensed into the maximization of the electric microwave field per photon at the atom position $z_0 = \qty{80}{\micro\meter}$ above the capacitor gap.
The strength of the interaction on the single-quantum level is encoded in the single-photon coupling rate~$g$, which enters the interaction Hamiltonian of the Rabi model as
\begin{equation}
  \hat H_{\mathrm{int}} / \hbar = g (\hat a + \hat a^\dagger) (\hat\sigma_+ + \hat\sigma_-) .
\end{equation}
Here, $\hat a$ and~$\hat a^\dagger$ are the annihilation and creation operators of the microwave field mode, respectively, and $\hat\sigma_+ = \ket{r_2} \bra{r_1}$ and $\hat\sigma_- = \ket{r_1} \bra{r_2}$ are the raising and lowering operators of the Rydberg state pair.
The coupling rate~$g$ is closely related to the vacuum Rabi frequency $\Omega_0 = 2 g$ (for zero detuning between atom and photon) and describes the rate of energy exchange between the two systems.
In terms of practical variables, $g$~can be expressed as
\begin{equation}
  g = \frac{E_{\mathrm{zpf}} d_0}{\hbar}
\end{equation}
with the zero-point fluctuation amplitude of the electric microwave field at the position of the atom $E_{\mathrm{zpf}}$ and the transition dipole moment of the Rydberg state pair~$d_0$, cf.~again Fig.~\subref{fig:design-concept}{d}.

Since the atoms are located considerably above the chip surface, it is not sufficient to maximize the zero-point voltage~$V_{\mathrm{zpf}} = \sqrt{\hbar \omega_0 / 2 C}$ of the microwave circuit as it is often the case in planar circuit QED~\cite{Wallraff2004, Bosman2017}.
If that were the case we would just have to minimize the capacitance~$C$ while keeping the resonance frequency at \qty{11}{\giga\hertz}.
Instead, we need to numerically scan a large parameter space of circuit parameters in order to find the optimum regime.
An overview of the relevant variables and an illustration of the optimization procedure are presented in Fig.~\ref{fig:parameter-optimization}.
Since we have fixed the length of the capacitor plate to $l = \qty{1}{\milli\meter}$, the remaining parameters are the width of the plate~$a$, the size of the capacitor gap to ground~$b$ and the length of the inductance~$s$.
The length of the inductance~$s$ is mainly a control knob to keep the resonance frequency at $\omega_0 = 2 \pi \times \qty{11}{\giga\hertz}$ for all combinations of $a$ and~$b$, but it also contributes to unwanted stray capacitances.

The procedure to determine $E_{\mathrm{zpf}}$ for each set of parameters $(a, b)$ is as follows:
First, we need to determine the value~$s$ that results in the desired resonance frequency.
Initially, we use an iterative procedure of simulating the circuit with the software package \software{Sonnet} for different values of~$s$ until the resulting resonance frequency is~$\qty{11}{\giga\hertz}$.
Once the function $(a, b) \mapsto s$ is known for a few parameter sets, we can estimate its value at further points by interpolation and in this way quickly scan a large parameter space.
Results of this process can be seen in Fig.~\subref{fig:parameter-optimization}{b}.
Next, we determine the capacitance~$C$ for each parameter set $(a, b, s)$ by numerical simulation with the software package \software{Comsol Multiphysics} in combination with an analytical correction which takes the mode shape of the microwave fields into account.
For completeness, we calculate the corresponding inductance $L = 1 / \omega_0^2 C$, cf.~Fig.~\subref{fig:parameter-optimization}{c}.
Details regarding the simulations and the capacitance correction can be found in Appendices~\ref{sec:app_simudetails} and \ref{sec:app_C_correction}.

From the capacitance simulations with \software{Comsol Multiphysics}, we furthermore extract the spatial field distribution $\vec E_{\mathrm{dc}}(\vec r)$ for $V_{\mathrm{dc}} = \qty{1}{\volt}$, cf.~Fig.~\subref{fig:design-concept}{c}, which we translate into the microwave zero-point field amplitude by
\begin{equation}
  E_{\mathrm{zpf}}(\vec r) = \sqrt{\frac{\hbar \omega_0}{2 C}} \frac{\abs{\vec E_{\mathrm{dc}}(\vec r)}}{V_{\mathrm{dc}}} .
\end{equation}
We now need to find the optimal range for $a$ and~$b$ which maximizes $g \propto E_{\mathrm{zpf}}$ at $\vec r = (0, 0, z_0)$.
Luckily, we find in these simulations that the optimum range of~$b$ is only slightly dependent on~$a$ and vice versa, and that $a$ in general has not a very strong impact on~$g$ in a broad interval of reasonable widths, since the electric field will be concentrated near the gap to ground anyways.
The optimum ranges for both variables are comfortably large, with deviations in either of them as large as~\qty{50}{\percent} reducing the coupling rate only slightly, cf.~Figs.~\subref{fig:parameter-optimization}{d, e}.
This relaxes the requirements for the exact design values and the fabrication precision.
We find $(a, b, s) = \qtylist[list-units=bracket, list-final-separator={, }]{60; 50; 1390}{\micro\meter}$ to be the optimal parameters, resulting in $C = \qty{136}{\femto\farad}$, $L = \qty{1.53}{\nano\henry}$ and a coupling strength $g_{\mathrm{max}} = 2 \pi \times \qty{433}{\kilo\hertz}$, cf.~Fig.~\subref{fig:parameter-optimization}{e}, orders of magnitude larger than in any previous chip-based implementation~\cite{hattermann2017, kaiser2022}.

Such a large value of~$g$ is highly promising for achieving the strong-coupling regime, which is defined by $2 g > \kappa, \gamma_{\mathrm{Ry}}$, where $\kappa$ is the total linewidth of the superconducting circuit and $\gamma_{\mathrm{Ry}} / 2 \pi \sim \qtyrange[range-units=bracket]{1}{10}{\kilo\hertz}$ is the atom decay rate of the Rydberg state pair under consideration here.
Optimized superconducting coplanar waveguide cavities have been demonstrated to allow for single-photon quality factors $Q_{\mathrm{int}} > \num{e6}$~\cite{bruno2015}, which translates to $\kappa_{\mathrm{int}} < 2 \pi \times \qty{11}{\kilo\hertz}$ and hence to $2 g / \kappa \sim 80$ for our $\omega_0$.
In our setting, however, the cavity layout is highly specialized and not an ideal coplanar waveguide, the chip is not packaged compactly in a radiation-tight housing and there will always be some infrared or optical stray light hitting the superconducting film.
It is therefore important to characterize as a next step a physical implementation of a typical device.

\subsection{Experimental device and setup}

\begin{figure*}
  \includegraphics{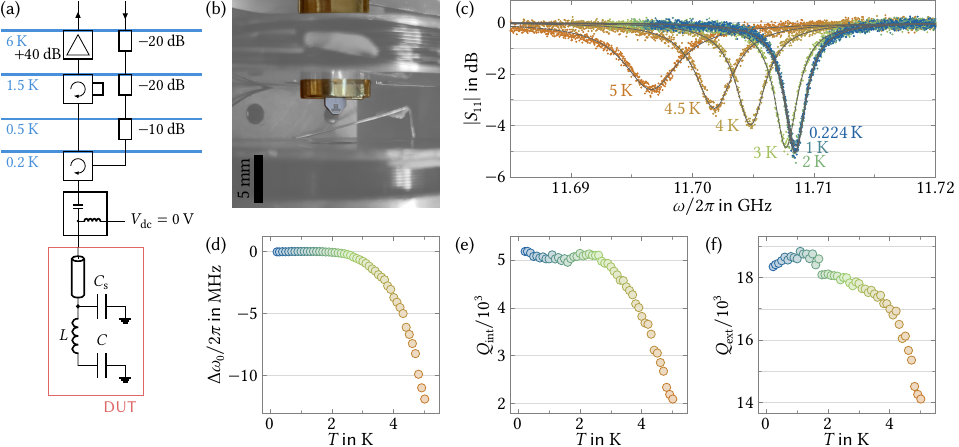}
  \vskip -6pt\relax
  \titlecaption{Cavity characterization in a cold-atom dilution refrigerator}{%
    \sublabel{a}~Schematic of the microwave-part of the dilution fridge setup.
    A cascade of attenuators~(boxes) is mounted to the input line (little arrows at the top indicate the microwave signal direction) to reduce the thermal-noise level of the incoming microwave signal to sub-kelvin levels.
    The signal is then routed to the device under test~(DUT, cf.~Fig.~\ref{fig:design-concept}) via a cryogenic circulator and through a bias-tee.
    The bias-tee combines the microwave signal with an applied dc~voltage directly in front of the cavity, but none is applied during the measurements presented in this figure.
    The reflected signal is amplified by \qty{\sim 40}{\deci\bel} using a cryogenic HEMT amplifier on the \qty{6}{\kelvin}~stage.
    A cryogenic isolator on the \qty{1.5}{\kelvin}~plate is used to shield the superconducting resonator from thermal noise emitted by the amplifier.
    \sublabel{b}~Photograph of the mounted resonator.
    The chip-tip is uncovered in order to allow access for lasers and atoms.
    Above and below the resonator two superconducting coils can be seen that make up the magnetic quadrupole trap used to load the atoms into the optical tweezers (coils not used here).
    \sublabel{c}~Reflection~$\abs{S_{11}}$ of the microwave resonator at different temperatures between \qty{224}{\milli\kelvin} and \qty{5}{\kelvin}.
    Colored dots are experimental data, gray lines are corresponding fit curves, cf.~Appendix~\ref{sec:app_fitting}.
    The curves for $T \leq \qty{2}{\kelvin}$ lie on top of each other.
    From the fit lines, we extract the cavity resonance frequency~$\omega_0$ and the internal and external quality factors $Q_{\mathrm{int}} = \omega_0 / \kappa_{\mathrm{int}}$ and $Q_{\mathrm{ext}} = \omega_0 / \kappa_{\mathrm{ext}}$.
    \sublabel{d},~\sublabel{e},~\sublabel{f}~Key fit parameters versus temperature~$T$.
    We show the resonance frequency shift $\Delta\omega_0 = \omega_0(T) - \omega_0(T_{\mathrm{min}} = \qty{0.2}{\kelvin})$ as well as the internal and external quality factors $Q_{\mathrm{int}}$ and~$Q_{\mathrm{ext}}$.
    All parameters show a nearly constant value for $T \lesssim \qty{2.5}{\kelvin}$, and a considerable decrease with increasing temperature for larger $T$.
    We attribute the discontinuity in~$Q_{\mathrm{ext}}$ at around~\qty{1.7}{\kelvin} to aluminum coaxial cables and/or aluminum wire-bonds becoming superconducting.
    The decrease of $\omega_0$ and~$Q_{\mathrm{int}}$ with increasing~$T$ can be explained by thermal quasiparticles.
  }
  \label{fig:T-characterization}
\end{figure*}

We have fabricated and characterized a superconducting cavity chip with parameters very close to the optimal ones determined in the previous section.
As geometrical parameters we chose an inductance wire length $s = \qty{1000}{\micro\meter}$, capacitor dimensions $a = \qty{120}{\micro\meter}$ and $b = \qty{40}{\micro\meter}$, and as plate length we stick to the fixed value $l = \qty{1000}{\micro\meter}$.
This results in a design frequency $\omega_0 = 2\pi\times \qty{11.6}{\giga\hertz}$ ($C = \qty{160}{\femto\farad}$, $L = \qty{1.18}{\nano\henry}$), i.e.~a slightly higher resonance frequency than the previously discussed~\qty{11}{\giga\hertz} which is a precaution for the case that we cannot compensate for all parasitic dc fields with~$V_{\mathrm{dc}}$ and are therefore restricted to the upper range of possible electric fields $E_{\mathrm{dc}} \gtrsim \qty{1}{\volt\per\centi\meter}$, cf.~Fig.~\subref{fig:design-concept}{d} and Ref.~\cite{kaiser2022}.
The higher resonance frequency results in an expected coupling rate $g = 2 \pi \times \qty{438}{\kilo\hertz}$, slightly larger than the value in Fig.~\ref{fig:parameter-optimization}.

The circuit consists of three micro-patterned thin-film layers on top of a tip-tapered sapphire substrate.
The first layer is superconducting niobium with a thickness of \qty{80}{\nano\meter} and defines all superconducting structures except for the top-plate of the shunt capacitor.
The second layer is \qty{70}{\nano\meter} thick \ce{Si3N4} as dielectric for the shunt capacitor and the third layer is again superconducting niobium of \qty{120}{\nano\meter} thickness to finish the parallel-plate capacitor~$C_{\mathrm s}$.
The shunt capacitance $C_{\mathrm s} \approx \qty{20}{\pico\farad}$ is designed for a high external quality factor $Q_{\mathrm{ext}} = \omega_0 Z_0 C_{\mathrm s} (C + C_{\mathrm s}) / C \sim 10^5$ and in contrast to similar earlier implementations~\cite{bosman2015, schmidt2020, uhl2024} the top plate is galvanically connected to the ground planes in order to minimize the impact of a potential shunt inductance.
More details regarding the chip fabrication can be found in Appendix~\ref{sec:app_devicefab}.

For its characterization, the \qtyproduct[product-units=power]{10 x 3.5}{\milli\meter} large superconducting chip is mounted in a cold-atom-compatible way into a suitable dilution refrigerator, cf.~Fig.~\ref{fig:T-characterization}.
The tapered tip of the chip including the complete cavity protrudes over the edge of its microwave housing by \qty{2}{\milli\meter}.
Inside the gold-plated copper housing, the chip is wire-bonded directly to the housing along the two long chip edges for grounding and to a Rogers microwave printed circuit board~(PCB) at the remaining short edge with the CPW feedline launcher, cf.~Appendix~\ref{sec:app_devicefab}.
The PCB connects the chip to a surface-mounted SMP connector by means of another CPW transmission line.
Then, the complete package is inserted through the circular bore of a small superconducting magnet coil which is part of a magnetic quadrupole trap for cold atoms.
By design and careful mounting, everything is aligned in a way that the microwave capacitor~$C$ on the chip is located close to the center of the quadrupole coils, where the atoms arrive from their \qty{6}{\kelvin}~preparation stage and where they are transferred from the magnetic into the optical dipole trap.
Simultaneously, the magnet-center is positioned directly on the axis of optical access and the chip surface is oriented parallel to that axis.

The microwave input and output lines in the fridge are equipped with multiple high-frequency components which are typical for the characterization of superconducting quantum circuits at \unit{\milli\kelvin}~temperatures, cf.~Fig.~\ref{fig:T-characterization} and Appendix~\ref{sec:app_setup}.
A minor difference to common \unit{\milli\kelvin}~setups is that all the temperature stages in our cold-atom fridge have somewhat elevated temperatures compared to standard systems~\cite{jessen2014}, and so the lowest temperature we can (currently) achieve is $T_{\mathrm{min}} \gtrsim \qty{200}{\milli\kelvin}$.
On the input line, we have mounted several discrete attenuators and highly attenuating stainless steel coaxial cables (total input attenuation \qty{\sim 70}{\decibel} at~\qty{11}{\giga\hertz}) to suppress the noise level of the incoming signals to an equivalent temperature close to~$T_{\mathrm{min}}$.
On the mixing chamber and just before the signals reach the superconducting chip, we have installed a cryogenic circulator and a bias-tee to enable a reflection measurement in combination with the application of a dc~voltage.
The reflected microwave signal is routed with partially superconducting coaxial cables (up to the $T = \qty{0.5}{\kelvin}$ plate) through the circulator towards a cryogenic high-electron-mobility-transistor~(HEMT) amplifier in the output line.
To additionally shield the device from thermal noise emitted by the HEMT, a cryogenic isolator is inserted at $T = \qty{1.5}{\kelvin}$ in between the circulator and the amplifier.

\subsection{Measured cavity parameters vs \texorpdfstring{$T$}{T}}
The spectroscopic cavity characterization is performed by means of a vector network analyzer~(VNA), and we track the reflection response~$S_{11}$ around the circuit resonance frequency during the device cooldown to~\unit{\milli\kelvin}.
This way we are able to cover a large temperature span up to $T > \qty{4}{\kelvin}$, which could not be achieved as easily by the heater on the \unit{\milli\kelvin}~plate alone.
A separate temperature sensor is directly attached to the chip mounting-bracket to guarantee a minimal temperature-difference between sensor and sample.
The measured absorption resonances, cf.~Fig.~\subref{fig:T-characterization}{c} for a small selection of temperatures, are analyzed by a data fitting procedure from which we obtain the resonance frequency~$\omega_0$ and the internal and external quality factors $Q_{\mathrm{int}}$ and~$Q_{\mathrm{ext}}$, cf.~Appendix~\ref{sec:app_fitting}.
At the lowest available temperature $T_{\mathrm{min}} = \qty{224}{\milli\kelvin}$, we find $\omega_0 = 2 \pi \times \qty{11.708}{\giga\hertz}$, $Q_{\mathrm{int}} = \num{5.2e3}$ and $Q_{\mathrm{ext}} = \num{18.3e3}$, indicating that the resonator is in the undercoupled regime.
Up to \qty{\sim 2.5}{\kelvin} the cavity properties remain nearly constant, and start to decrease for further increasing temperatures.
A decrease of $\omega_0$ and~$Q_{\mathrm{int}}$ with increasing temperature can be attributed to thermal quasiparticles in the superconductor and a corresponding increase of kinetic inductance.
The slight decrease of $Q_{\mathrm{ext}}$ with increasing~$T$ on the other hand could either be related to a frequency-dependent impedance of the microwave circuitry connected to the chip (cable resonances) or to a non-negligible contribution of $Q_{\mathrm{int}}$ to~$Q_{\mathrm{ext}}$ based on interferences (e.g.~in the circulator) and incorrectly attributed losses during the fitting procedure~\cite{rieger2023}.
These factors might also be responsible for the discontinuity of $Q_{\mathrm{ext}}$ at $T \sim \qty{1.7}{\kelvin}$ which occurs approximately when the aluminum coaxial cables and wire bonds in the setup become superconducting.

Overall, the cavity shows very promising characteristics for the envisioned hybrid system.
The resonance frequency is close to its design value and in an ideal range for the targeted Rydberg transition.
The external quality factor is smaller than expected, but in principle allows $\kappa = \omega_0 / Q \sim 2 \pi \times \qty{600}{\kilo\hertz}$ and can be adjusted in future implementations, e.g.~by an increased plate area of~$C_{\mathrm s}$, a thinner layer of~\ce{Si3N4} or a material with a higher~$\epsilon_{\mathrm r}$ such as amorphous silicon.
For the present device and first hybrid experiments, however, $Q_{\mathrm{ext}}$~is well-suited, since it does not dominate the total linewidth while still allowing for a comfortable cavity characterization by a considerable resonance-dip depth of~\qty{5}{\decibel}.
The internal quality factor $Q_{\mathrm{int}} = \num{5.2e3}$ finally is lower than in many other implementations of superconducting circuits, and currently we are not certain why.
Possible reasons include stray (infrared) light reaching the superconductor through the cryostat windows, dipole-radiation losses due to a lack of metal housing, as well as surface, interface and dielectric losses on the chip itself.
The latter seems improbable because we can achieve $Q_{\mathrm{int}} > \num{e5}$ at \unit{\milli\kelvin}~temperatures with our fabrication recipe but different cavity layouts and proper microwave-packaging.
From our estimates of the radiation losses, we would also expect an at least tenfold higher~$Q_{\mathrm{int}}$.
However, if radiation turns out to be the culprit, a careful design of the free-space density of states around the chip tip, e.g.~by metal plates that do not impede the optical path, is likely to suppress it to a satisfactorily low level.
Optical and infrared stray light could be minimized by suitable filter-shielding of the windows for optical access.

We plan to investigate the exact loss mechanisms of our cavity in dedicated experiments in the future and we are optimistic that we can improve the quality factor by at least one order of magnitude doing so.
Nevertheless, the cavity does already now facilitate getting close to the strong-coupling regime with $\Omega_0 \approx \kappa / 3$ which would exceed all previous on-chip realizations by orders of magnitude~\cite{hattermann2017, kaiser2022}.
Since our setup involves not only a single Rydberg atom but potentially many of them simultaneously (in an ensemble of \numrange{e4}{e6} atoms), we also expect to observe a transition from the weak-coupling to the strong-coupling regime with increasing number of Rydberg atoms.
The Dicke model predicts a scaling of the collectively enhanced vacuum Rabi frequency with the root of the number of Rydberg atoms $\smash{\sqrt{N_{\mathrm{Ry}}}}$~\cite{dicke1954}, which can in turn be chosen by the intensity of the laser pulses that excite the atoms from the electronic ground state to the Rydberg manifold, cf.~Fig.~\ref{fig:design-concept}.
In order to achieve $\Omega_N = \smash{\sqrt{N_{\mathrm{Ry}}}} \, \Omega_0 > \kappa$, we need $N_{\mathrm{Ry}} \gtrsim 10$, which is feasible for us considering a realistic Rydberg blockade radius on the order of~\qty{10}{\micro\meter}~\cite{saffman2010, stecker2020}.
Further optimization strategies for considerably enhanced vacuum Rabi frequencies of a single atom are explained in \hyperref[sec:discussion]{the discussion section}.

\subsection{Measured cavity parameters vs \texorpdfstring{$V_{\mathrm{dc}}$}{Vdc}}

\begin{figure}[b]
  \includegraphics{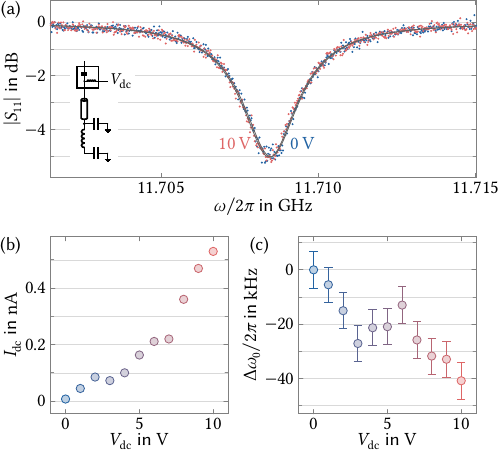}
  \vskip -5pt\relax
  \titlecaption{Cavity characterization with applied bias-voltage~$V_{\mathrm{dc}}$}{%
    \sublabel{a}~Reflection~$\abs{S_{11}}$ of the microwave resonator at $T = \qty{0.2}{\kelvin}$ with $V_{\mathrm{dc}} = \qty{0}{\volt}$ and $V_{\mathrm{dc}} = \qty{10}{\volt}$ in direct comparison.
    Dots are experimental data and gray lines are overlayed fit curves.
    No significant shift or widening of the resonance dip is discernible, indicating the robustness of the cavity against dc~voltages up to at least~\qty{10}{\volt}.
    \sublabel{b}~Leakage current~$I_{\mathrm{dc}}$ recorded during measurements with different~$V_{\mathrm{dc}}$.
    The observed resistance of \qty{\sim 20}{\giga\ohm} is in the range expected for leakage through the \ce{Si3N4} shunt capacitor~\cite{bosman2015}.
    \sublabel{c}~Resonance frequency shift $\Delta\omega_0$ versus applied dc~voltage.
    Error bars indicate the estimated standard errors as given by the fit routine.
  }
  \label{fig:Vdc-characterization}
\end{figure}

As a second important device characterization we apply a dc voltage~$V_{\mathrm{dc}}$ to the superconducting resonator, which will later be important to compensate for parasitic dc~electric fields from e.g.~surface adsorbates and to tune the Rydberg transition frequency into (and out of) resonance with the cavity.
Hence, we apply a variable~$V_{\mathrm{dc}}$ to the dc~line of the bias-tee, monitor a possible leakage current~$I_{\mathrm{dc}}$ and measure the cavity reflection~$S_{11}$ for each static value of~$V_{\mathrm{dc}}$.
The corresponding results are summarized in Fig.~\ref{fig:Vdc-characterization}.
For voltages up to $V_{\mathrm{dc}} = \qty{10}{\volt}$, the cavity lineshape and resonance frequency remain unmodified to the naked eye.
We observe, however, a small leakage current of up to~\qty{\sim 500}{\pico\ampere} which corresponds to a resistance of~\qty{20}{\giga\ohm}.
Such a resistance is consistent with earlier observations in analogous systems with trilayer parallel-plate capacitors made of \ce{Si3N4}~\cite{bosman2015}, but we also cannot exclude a contribution from leakage currents through the bias-tee or the dc~wires and connectors.

From fits to the cavity resonance, we find a small shift of the resonance frequency with increasing voltage, cf.~Fig.~\subref{fig:Vdc-characterization}{c}, and constant values of $Q_{\mathrm{int}}$ and~$Q_{\mathrm{ext}}$ (not shown).
However, due to the smallness of the frequency shift, which is comparable to the uncertainty of its absolute value, it is not completely clear whether it is an actual change of the resonance frequency or an artifact e.g.~due to changed properties of the bias-tee.
On the one hand, a red-shift would for instance be compatible with an electrostatic compression of the dielectric in the shunt capacitor; on the other hand, this should also lead to a change of~$Q_{\mathrm{ext}}$.
A future cavity with a much higher~$Q$ could provide more insight into this effect.
In essence though, the cavity is very robust with respect to an applied dc~voltage up to at least~\qty{10}{\volt}, which we expect to be more than sufficient considering that this corresponds to an dc~electric field of $E_{\mathrm{dc}} \approx \qty{370}{\volt\per\centi\meter}$ at the position of the atoms.
Typical adsorbate fields are on the order of~\qty{10}{\volt\per\centi\meter}~\cite{hattermann2012, kaiser2022, ocola2024} and the Rydberg transition is optimally tuned into resonance at \qty{\lesssim 2}{\volt\per\cm}, cf.~Fig.~\ref{fig:design-concept}.
We finally note that we have applied even higher voltages to other chips at~\qty{4.2}{\kelvin} which revealed that the breakdown voltage of the shunt capacitor seems to be around $V_{\mathrm{dc}} \approx \qty{40}{\volt}$.

\subsection{A concept for maximal field homogeneity}
\begin{figure}
  \includegraphics{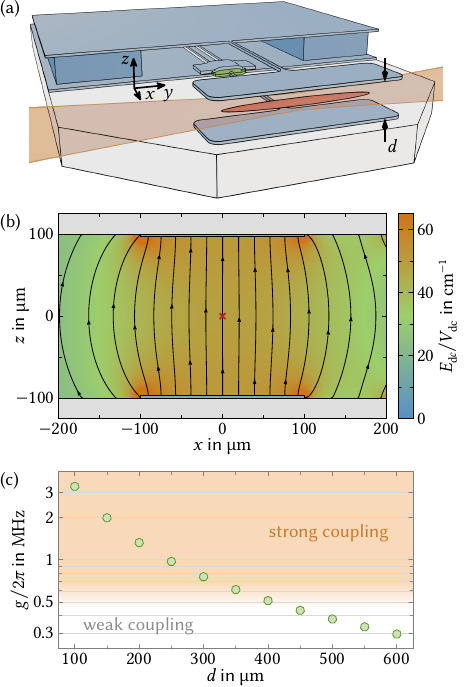}
  \vskip -5pt\relax
  \titlecaption{Modified cavity layout for improved field homogeneity and coupling strength}{%
    \sublabel{a}~Schematic of an alternative design consisting of two chips flipped on top of each other.
    The bottom chip is similar to the design presented in Fig.~\ref{fig:design-concept}, except that there is no ground plane close to the capacitance plate.
    The top chip (substrate not shown for clarity) features an identical capacitance plate which is connected by a mirror-inductance to a large ground plane.
    Two cuboid pieces of niobium control the distance~$d$ between the two chips while simultaneously acting as superconducting vias connecting the ground planes (other materials are also under consideration).
    \sublabel{b}~Cross-sectional schematic at the position of the atomic ensemble and the numerically simulated ratio of dc~electric field strength~$E_{\mathrm{dc}}$ over applied voltage~$V_{\mathrm{dc}}$ versus position for $d = \qty{200}{\micro\meter}$, $a = \qty{200}{\micro\meter}$ and $l = \qty{400}{\micro\meter}$; electric field lines are indicated.
    Values close to the plate edges that exceed the scale of the colorbar are drawn in the orange of the maximum.
    Thanks to the parallel plate capacitor geometry, the field around the atomic ensemble position (marked~\plotref{plt:cloudpos}\,) is very homogeneous.
    \sublabel{c}~Expected coupling strength~$g$ for different chip distances~$d$ (other parameters are scaled with~$d$, see main text) and a resonance frequency of $\omega_0 = 2 \pi \times \qty{11}{\giga\hertz}$.
    Assuming a resonator quality factor around~\num{10000}, we reach the strong coupling regime for plate distances below~\qty{350}{\micro\meter}, and for the smallest investigated distance $d_{\mathrm{min}} = \qty{100}{\micro\meter}$ we get $2 g_{\mathrm{max}} = 2 \pi \times \qty{6.6}{\mega\hertz} = 6 \kappa$.
  }
  \label{fig:3d-design-concept}
\end{figure}

Before we conclude our report, we will present a modified concept for the superconducting cavity as a perspective possibility which comes with considerable advantages, but at the expense of a higher complexity.
The basic idea is presented in Fig.~\subref{fig:3d-design-concept}{a} and is not based on a single superconducting chip but on two nearly identical chips facing each other with a distance~$d$ in a flip-chip assembly.
Each of the two chips contains an inductive wire and one half of a (vacuum) parallel plate capacitor, inside of which the dipole trap with the cold atoms will be located.
As a spacer between the chips, a superconducting foil, low-loss dielectric patches or indium bump bonds might be suitable choices.
The bottom chip furthermore contains the microwave feedline and the shunt capacitor as did the 2D~implementation discussed above.
The major advantages of this approach lie in the symmetry and in the spatial homogeneity of the electric dc and microwave fields.

For a completely symmetric set of chips, the adsorbate fields from the two chips might either cancel in the center of the capacitor or combine to a homogeneous field as well, which can then easily be compensated by an applied field~$E_{\mathrm{dc}}$.
Even if the parasitic fields remain inhomogeneous or are accompanied by a considerable gradient, it will be possible to relocate the atoms in the capacitor volume to a more favorable position by shifting the laser axis, affecting only the adsorbate fields with the shift.
The property of position-invariance is also very advantageous in case there are any vibrations of the superconducting chips with respect to the dipole trap, e.g.~induced by the refrigerator pulse-tube.
In principle, vibrations could be compensated by a real-time feedback to the dipole-trap position, however the experiment would be much simpler, and most likely more stable, if vibrations did not matter at all, just as it is the case for homogeneous capacitor fields.
Finally, homogeneous fields will minimize any inhomogeneous broadening, making coherent state transfer between the superconducting circuit and the atoms much easier.

Beyond the aspect of increased homogeneity, the~3D approach may also lead to higher vacuum Rabi frequencies than the planar version.
To demonstrate the possibility to reach the strong-coupling regime without a strict parameter optimization, we assume that the total circuit capacitance is defined by the two parallel plates and calculate the corresponding dc~capacitance~$C_{\mathrm{3D}}$ including the two substrates by means of \software{Comsol Multiphysics}.
Just as in the 2D~case, we also extract the field~$\vec E_{\mathrm{dc}}(\vec r)$ for an applied voltage $V_{\mathrm{dc}} = \qty{1}{\volt}$ and calculate~$g$ for a resonance frequency $\omega_0 = 2 \pi \times \qty{11}{\giga\hertz}$ as before.

In order to achieve maximum coupling strength, the capacitor dimensions should be chosen just large enough to ensure field homogeneity in the region of the atom cloud.
To that end, we choose $a = d$ and $l = \min(2 d, \qty{250}{\micro\meter})$ for a given chip distance~$d$.
The smallest chip distance ensuring only a negligible amount of laser power directly reaching the two chips is~\qty{\sim 100}{\micro\meter}.
Since such a small laser-chip distance might not be practically feasible, we also consider chip distances up to~\qty{600}{\micro\meter} in our simulations.
See Appendix~\ref{sec:app_3D_parameters} for a more detailed discussion of these parameter considerations.

The resulting simulated coupling rates are plotted in Fig.~\subref{fig:3d-design-concept}{c}.
Notably, the transition to the strong coupling regime $2 g > \kappa$ is observed for distances $d \lesssim \qty{350}{\micro\meter}$ if we assume a realistic cavity quality factor $Q \sim \num{10000}$.
For the smallest possible distance $d_{\mathrm{min}} = \qty{100}{\micro\meter}$ we even get $g_{\mathrm{max}} = 2 \pi \times \qty{3.3}{\mega\hertz}$, which corresponds to $2 g = 6 \kappa$ in the case of $Q = \num{10000}$ or $2 g = \kappa$ in the case $Q = \num{1656}$.

An intuitive measure for the field homogeneity, taking the atom distribution into account, is the weighted standard deviation of the relative difference from the field strength~$E_0$ in the center of the cloud
\begin{equation}
  \eta = \sqrt{ \int_{\mathbb R^3} \rho_{\mathrm{sp}}(\vec r) \, \Bigl( \frac{E(\vec r)}{E_0} - 1 \Bigr)^2 \dif{\vec r} } .
\end{equation}
Since contributions outside the cloud region defined by $d_{\ce{Rb}}$ and~$l_{\ce{Rb}}$ are suppressed by~$\rho_{\mathrm{sp}}$, it is sufficient to consider the integral in that region.
Evaluating the expression numerically using the field strength $E = \abs{\vec E_{\mathrm{dc}}}$ from the \software{Comsol Multiphysics} simulations, we obtain $\eta_{\mathrm{2D}} = \qty{0.5}{\percent}$ for the 2D~implementation discussed above.
The 3D~layout with $d = \qty{450}{\micro\meter}$ has a comparable coupling strength and for this we find $\eta_{\mathrm{3D}, 450} = \qty{0.02}{\percent}$, an improvement by a factor of~\num{25}.
While it increases the coupling strength, scaling the layout to smaller values of~$d$ also causes the homogeneous region to shrink and thus increases the inhomogeneity across the atom cloud.
For $d = \qty{200}{\micro\meter}$ we find $\eta_{\mathrm{3D}, 200} = \qty{0.2}{\percent}$, a somewhat less impressive improvement over the 2D~implementation, but with a threefold higher coupling rate.
This value can be further improved by optimizing the capacitor shape, but this may not be necessary in the end, as inhomogeneities become less relevant further into the strong coupling regime~\cite{diniz2011}.

We note that both chips in the flip-chip assembly could be supplemented with a ground plane surrounding the capacitor plates in a sufficiently large distance such that the dc and microwave electric fields are still concentrated in the space between the two rectangular plates on the two chips.
We omitted this potential ground plane here for the sake of simplicity.
If included, we expect it to somewhat reduce the field homogeneity and the vacuum Rabi frequency, but at the same time lead to a suppression of potential dipole-radiation losses into free space by counter-charges.
The optimal configuration remains to be investigated in the future.

\section{Discussion}\label{sec:discussion}
In this work we have presented the engineering and optimization of a superconducting microwave chip for the tunable coupling to optically trapped ultra-cold Rydberg atoms in a dilution refrigerator.
We focused on a suitable chip outline, an optimized laser-chip distance and on the ideal circuit parameter-set for maximized interaction rates.
Despite conservative estimates of various parameters such as the atomic ensemble temperature and a moderate cavity quality factor in the experimental implementation of the superconducting resonator of $Q_{\mathrm{int}} \sim \num{5200}$, our device allows for a single-atom vacuum Rabi frequency in proximity to the strong-coupling regime with $\Omega_0 \sim \kappa / 3$.
Since we intend to couple to an atomic ensemble, however, we expect a single-photon coupling-enhancement by the root of the number of atoms excited to the Rydberg manifold by the laser~$\smash{\sqrt{N_{\mathrm{Ry}}}}$.
Hence, for $N_{\mathrm{Ry}} > 10$ we reach the strong coupling region.
For future devices we furthermore believe to be able to increase~$Q_{\mathrm{int}}$ by at least one order of magnitude.
Finally, we have presented a perspective flip-chip cavity layout which could lead to enhanced field homogeneity and potentially larger coupling rates.

Even the planar implementation could be re-designed in the future to allow for further increased vacuum Rabi frequencies if it turns out that the atom characteristics permit it.
If the atomic ensemble in the dipole trap really can be cooled to $T \lesssim \qty{1}{\micro\kelvin}$ and if the surface adsorbate fields can be compensated sufficiently well with the applied dc~voltage, the entire circuit and chip could be implemented in a much smaller version.
The capacitance length could be reduced by a factor of five, and so could the chip width by a corresponding amount.
As a consequence, the laser could be focused to $w_{\mathrm{dp}} = \qty{10}{\micro\meter}$ and the trap-center could approach the chip to \qty{\sim 40}{\micro\meter}, which in turn allows for a smaller capacitor plate width and a smaller gap to ground.
Overall, these reductions alone would lead to an increase in~$E_{\mathrm{zpf}}$ by a factor of~\num{\sim 5}.
More sophisticated possibilities would include deep etching of the substrate to minimize the stray capacitance through the substrate or the use of a substrate material with much smaller~$\epsilon_{\mathrm r}$.
To compensate for the strongly reduced~$C$ by all these techniques, we could increase the inductance by using a spiral inductor or a very thin film with high kinetic inductance.

Our work uncovered an extensive overview over most of the critical aspects when interfacing superconducting microwave chips with optically trapped atoms and demonstrates a way to deal with the corresponding challenges in order to find a satisfying best-balance compromise.
The laid out strategies have the potential to bring dc-tunable and realistic superconductor-atom hybrid systems in the strong-coupling regime into experimental reality.
Once implemented, those systems can serve as foundations for novel quantum gates on superconducting microwave chips and towards unexplored regimes of light-matter interaction.
On the technological side it is promising for advances in quantum frequency conversion and for the implementation of atomic quantum memories on superconducting chips.

All raw data, measurement and data processing scripts as well as the numerical code used within this work to obtain the presented results will be made publicly available on the repository Zenodo with the identifier~\doi{10.5281/zenodo.14009235}.

\begin{acknowledgments}
  The authors thank Markus Turad, Ronny Löffler (instrument scientists at \lisaplus), Christoph Back and Florian Jessen for technical support, as well as Benedikt Ferdinand, Reinhardt Maier and Conny Glaser for helpful discussions.
  This research received funding from the Deutsche Forschungsgemeinschaft~(DFG) via grant numbers 421077991 (\mbox{KL~930/16-1}), 490939971 (\mbox{BO~6068/1-1}) and 465199066 (\mbox{FOR~5413/1}), as well as from the EU QuantERA project MOCA with the DFG grant number 491986552 (\mbox{FO~740/5-1} and \mbox{KL~930/18-1}).
  We also gratefully acknowledge support by the COST actions NANOCOHYBRI (CA16218) and SUPERQUMAP (CA21144).
\end{acknowledgments}

\appendix

\section{Stark map of the relevant Rydberg states}\label{sec:app_starkmap}
\begin{figure}
  \includegraphics{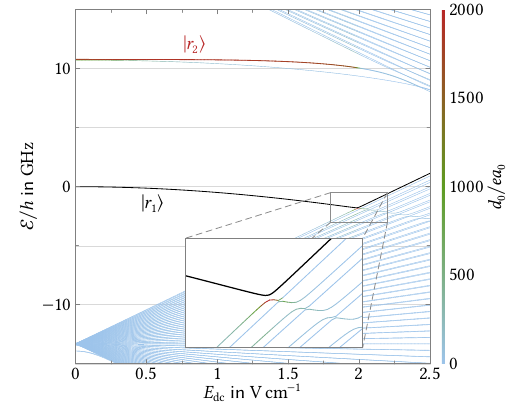}
  \titlecaption{Simulated Rydberg state energies in the vicinity of the targeted transition}{%
    The black line marks the initial state of the targeted high-$d_0$ transition $\ket{r_1} \to \ket{r_2}$, which at zero field corresponds to $\ket{\estate{59P_{3/2}}, m_j = \frac 32} \to \ket{\estate{58D_{5/2}}, m_j = \frac 32}$.
    The energy zero point is chosen to be $\mathcal E_1$ at~$E_{\mathrm{dc}} = 0$.
    For all states besides~$\ket{r_1}$, the color indicates the transition dipole moment~$d_0$ of the transition from~$\ket{r_1}$ to that state, cf.~Fig.\subref{fig:design-concept}{d}, with $\ket{r_2}$ appearing red due to the high associated transition dipole moment.
    At $E_{\mathrm{dc}} \approx \qty{2}{\volt\per\centi\meter}$, the energy of $\ket{r_1}$ bends upwards due to an avoided crossing with a lower energy state, shown magnified in the inset.
    This results in all transition frequencies $\mathcal E_{\mathrm tr} = \mathcal E - \mathcal E_1$ shown in Fig.~\subref{fig:design-concept}{d} bending down at that field strength.
  }
  \label{fig:stark-map}
\end{figure}

The (non-differential) Stark map of Rydberg state energies (calculated as described in Refs.~\cite{grimmel2015, kaiser2022}) in the vicinity of the states $\ket{r_1}$ and~$\ket{r_2}$ making up the target transition is shown in Fig.~\ref{fig:stark-map}.
The differential Stark map in Fig.~\subref{fig:design-concept}{d} results from this after subtracting~$\mathcal E_1$ from all lines.

We note that we omit all the possible transitions to negative relative frequencies in Fig.~\ref{fig:design-concept}, although they will also couple to the microwave cavity once they are resonant with it.
Their dipole transition matrix elements and corresponding coupling rates are several orders of magnitude smaller than that of the target transition, so they are negligible to first order, and their inclusion in Fig.~\subref{fig:design-concept}{d} would lead to an overcrowded image with considerably reduced clarity.

\section{Atom oscillation frequencies in a dipole trap}\label{sec:app_oscfreqs}
The radial and longitudinal~(axial) oscillation frequencies in the dipole trap $\omega_r$ and~$\omega_y$, respectively, determine the extension of the atomic cloud as described by Eq.~\eqref{eq:atom_density}.
We find the oscillation frequencies by revisiting the second-order Taylor expansion of a symmetric potential $U(x) = U(-x)$ around its minimum
\begin{equation}
  U(x) = U(0) + \frac 12 \del[2]x U(0) \times x^2 + \Olandau(x^4)
\end{equation}
and the differential equation of a particle with mass~$m$ in this potential (we neglect damping and external forces for the sake of simplicity)
\begin{equation}\label{eq:HO_EOM}
  m \ddot x + \del[2]x U(0) \times x = 0 .
\end{equation}
The solution of Eq.~\eqref{eq:HO_EOM} is a harmonic oscillation with (angular) frequency
\begin{equation}
  \omega_x = \sqrt{\frac{\del[2]x U(0)}{m}} .
\end{equation}
To find the frequencies for rubidium atoms in the dipole trap, we therefore need to calculate
\begin{align}
  \omega_r &= \sqrt{\frac{\del[2]r U_{\mathrm{dp}}(\vec 0)}{m_{\ce{Rb}}}} \\
  \omega_y &= \sqrt{\frac{\del[2]y U_{\mathrm{dp}}(\vec 0)}{m_{\ce{Rb}}}} .
\end{align}
The second partial derivatives of the Gaussian (radial) and the Lorentzian (axial) profile of the trap potential around their corresponding extrema are
\begin{align}
  \del[2]r \exp\Bigl( -\frac{2 r^2}{w_{\mathrm{dp}}^2} \Bigr) \evalwith[\bigg]{\mathrlap{r = 0}\hskip .8em} &= -\frac{4}{w_{\mathrm{dp}}^2} \\
  \del[2]y \Bigl( 1 + \frac{y^2}{l_{\mathrm R}^2} \Bigr)^{-1} \evalwith[\bigg]{\mathrlap{y = 0}\hskip .8em} &= -\frac{2}{l_{\mathrm R}^2} .
\end{align}
Combining everything and plugging in all relevant numbers, we find
\begingroup\abovedisplayskip\abovedisplayshortskip
\begin{align}
  \omega_r &= 2 \pi \times \qty{4.1}{\kilo\hertz} \\
  \omega_y &= 2 \pi \times \qty{50}{\hertz}
\end{align}
\endgroup
and with $m_{\ce{Rb}} = \qty{86.9}{\dalton}$ we arrive at the values for $\sigma_r$ and~$\sigma_y$ discussed in the context of Fig.~\ref{fig:dipole-trap}.

\section{Estimating the laser power reaching the chip}\label{sec:app_laser_on_chip}
Since optical photons can break Cooper pairs, destroy superconductivity and heat the mixing chamber of a dilution refrigerator, we estimate an upper threshold for the total optical power reaching the superconducting structures in our setup and adjust the laser-chip distance and the laser profile accordingly.
We consider two contributions and find that both are sufficiently small to be neglected when the parameters are chosen as discussed in the context of Fig.~\ref{fig:dipole-trap}.

The first way a photon from the dipole-trap laser can reach the chip is by directly hitting the superconductor in the Gaussian tail of the beam profile.
At the chip edge, i.e.\@ $y = l_{\mathrm{ch}} / 2 = \qty{0.8}{\milli\meter}$, the beam radius is $w_{\mathrm e} \approx \qty{20}{\micro\meter}$, meaning that the center of the beam is located at $z_0 = \qty{80}{\micro\meter} \approx 4 w_{\mathrm e}$ above the chip surface.
We therefore calculate the total power reaching the chip on or below its surface line and find
\begin{equation}\label{eq:P_dir}
  \begin{aligned}[b]
    P_{\mathrm{dir}}
    &= \sqrt{\frac 2\pi} \frac{P_{\mathrm{dp}}}{w_{\mathrm e}} \int_{z_0}^\infty \! \exp\Bigl( -\frac{2 \tilde z^2}{w_{\mathrm e}^2} \Bigr) \dif{\tilde z} \\
    &= \frac{P_{\mathrm{dp}}}{2} \biggl( 1 - \erf\Bigl( \! \sqrt 2 \frac{z_0}{w_{\mathrm e}} \Bigr) \biggr)
    = \qty{66}{\atto\watt}
  \end{aligned}
\end{equation}
for $P_{\mathrm{dp}} = \qty{50}{\milli\watt}$, which is clearly negligible~\cite{benevides2024, budoyo2016}.
Note that if the chip were not tapered but kept its full \qty{3.5}{\milli\meter} width at the position of the laser beam, while also maintaining $z_0 = \qty{80}{\micro\meter}$ and $w_{\mathrm{dp}} = \qty{15}{\micro\meter}$, the power incident on the superconductor would increase to~\qty{\sim 38}{\nano\watt}.

The second path a photon from the dipole laser can take is being off-resonantly scattered by the atoms in the trap.
The scattering rate of an atom at position~$\vec r$ in units of scattered photons per atom per second is given by~\cite{grimm2000}
\begin{equation}
  \def\pkern{\mkern -1mu\relax}
  \def\ekern{\mkern -4mu\relax}
  \def\okern{\mkern -2mu\relax}
  \def\rkern{\mkern -1mu\relax}
  \def\fkern{\mkern 2mu\relax}
  \begin{aligned}[b]
    \Gammasc(\vec r) \rkern&=\rkern \mkern -2mu\relax
    \biggl(
      \frac{\pi c_0^2}{2 \hbar \omega_1^3}
      \Bigl( \pkern \frac{\omega_{\mathrm{dp}}}{\omega_1} \pkern \Bigr)^{\ekern 3}
      \Bigl( \pkern \frac{\Gamma_1}{\omega_1 \okern-\okern \omega_{\mathrm{dp}}} \okern+\okern \frac{\Gamma_1}{\omega_1 \okern+\okern \omega_{\mathrm{dp}}} \pkern \Bigr)^{\ekern 2} \\
      &\phantom{ \biggl( } \mkern 8mu\relax +\rkern
      \frac{\pi c_0^2}{\hbar \omega_2^3}
      \Bigl( \pkern \frac{\omega_{\mathrm{dp}}}{\omega_2} \pkern \Bigr)^{\ekern 3}
      \Bigl( \pkern \frac{\Gamma_2}{\omega_2 \okern-\okern \omega_{\mathrm{dp}}} \okern+\okern \frac{\Gamma_2}{\omega_2 \okern+\okern \omega_{\mathrm{dp}}} \pkern \Bigr)^{\ekern 2}
    \biggr) \fkern I_{\mathrm{dp}}(\vec r) .
  \end{aligned}
\end{equation}
For a nearly constant intensity at the center of the trap $I_{\mathrm{dp}} = 2 P_{\mathrm{dp}} / \pi w_{\mathrm{dp}}^2$ we find $\Gammasc = 2 \pi \times \qty{15}{\per\second}$.
Assuming a total atom number of $N_{\mathrm{at}} = \num{e6}$ in the trap (though \num{e4} is more realistic), we obtain as total scattered power
\begin{equation}
  P_{\mathrm{sc}} = \hbar \omega_{\mathrm{dp}} \Gammasc N_{\mathrm{at}} \approx \qty{23}{\pico\watt} .
\end{equation}
This is again negligibly small -- even if we assume that all of it is absorbed by the superconductor.

\begin{figure}
  \includegraphics{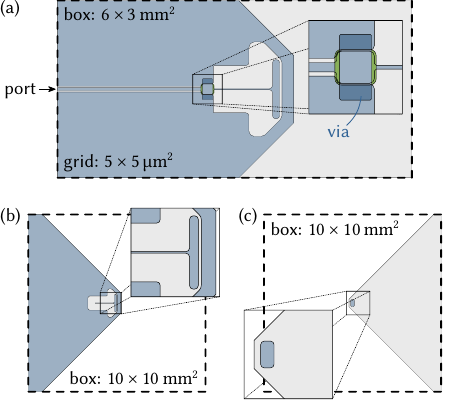}
  \titlecaption{Layouts used for rf and dc~simulations}{%
    Layout for \sublabel{a}~microwave simulation using \software{Sonnet} and \sublabel{b}~for dc~simulations using \software{Comsol Multiphysics}.
    \sublabel{c}~Analogous layout for the perspective 3D~design.
    The colors indicate the material, as in previous figures: Gray for sapphire, blue for niobium and green for silicon nitride.
    In addition, the vias allowing current flow between the bottom and the top layer of the shunt capacitor in \software{Sonnet} simulations are indicated by a darker blue.
    The simulation boxes are indicated as thick dashed lines and their dimensions are noted in the drawings.
    Insets show enlarged regions with small structures.
    The dimensions of the 2D~resonator shown in~\sublabel{b} correspond to the fields shown in Figs.~\subref{fig:design-concept}{c}, the ones of the 3D~resonator in~\sublabel{c} to those shown in Fig.~\subref{fig:3d-design-concept}{b}.
  }
  \label{fig:simu-layouts}
\end{figure}

\section{Details of simulations with \software{Sonnet} and \software{Comsol Multiphysics}}\label{sec:app_simudetails}
\software{Sonnet} simulations were performed using a \qtyproduct[product-units=power]{6 x 3 x 2}{\milli\meter} simulation box with the niobium film placed in its center, on top of a \qty{330}{\micro\meter} thick sapphire substrate ($\epsilon_{\mathrm r} = \num{10}$).
We choose the box somewhat smaller than the real device in order to keep simulation time manageable.
Due to a software limitation, the substrate fills the entire profile of the simulation box, i.e.~it does not follow the taper of the niobium layer.
The two niobium layers are modeled with a typical surface inductance value of $L_\square = \qty[per-mode=symbol]{0.13}{\pico\henry\per\sqr}$~\cite{surface-impedance-note}.
They are separated by a \qty{100}{\nano\meter} thick silicon nitride brick ($\epsilon_{\mathrm r} = \num{7.5}$) in the center and connected by lossles vias on either side.
A schematic of such a layout is shown in Fig.~\subref{fig:simu-layouts}{a}.
Since the smallest structure (the inductive wire) has a width of~\qty{20}{\micro\meter}, a grid size of~\qtyproduct[product-units=power]{5 x 5}{\micro\meter} is sufficient.

For the dc~field simulations using \software{Comsol Multiphysics} we only consider the components that determine the resonator capacitance.
That is, we omit the shunt capacitor, which affects the effective capacitance in principle but only with a small contribution, as well as the feedline.
We include the inductive wire and add additional spacing to ground around its back end, as shown in Fig.~\subref{fig:simu-layouts}{b}.
In order to minimize distortions of the field in the region of interest (boundary condition: no charges on the simulation box), we choose a large box of~\qtyproduct[product-units=power]{10 x 10 x 10}{\milli\meter}, with the niobium film placed in its center on top of a sapphire substrate, like in the \software{Sonnet} layout.
The chip is positioned in such a way that the capacitor gap, above which the atom cloud will be positioned, is located in the exact center of the box.
Since the total width of the chip was not determined yet and should not affect the relevant fields much, the chip taper is extended all the way to the box edges.
We triangulate the simulation mesh very finely (down to~\qty{1}{\micro\meter}) in the region of the capacitor and the atom cloud, and more loosely (up to \qty{500}{\micro\meter}) far away from the capacitor, where field gradients are low.

The layout for the dc~simulations for the perspective 3D~design is essentially the same, except that the ground plane and the inductive wire are omitted, leaving only the capacitor plate, and a second, identical chip is placed above, facing the first.
One such chip layout can be seen in Fig.~\subref{fig:simu-layouts}{c}.
The center between the two plates, where the atom cloud will be positioned, is located in the center of the simulation box, with the two chips symmetrically below and above it.

\section{Standing-wave correction of the circuit capacitance}\label{sec:app_C_correction}
\begin{figure}
  \includegraphics{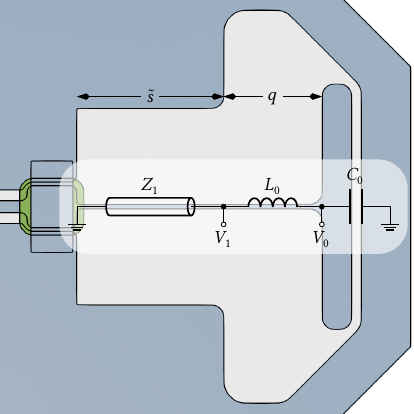}
  \titlecaption{Model for the standing-wave correction of the circuit capacitance}{%
    We model the leftmost inductive wire section with length $\tilde s = s - q$ as a CPW with characteristic impedance~$Z_1$ and the remaining part of length~$q$ as a lumped-element inductor~$L_0$.
    With the capacitance contribution~$C_0$ of the gap between the capacitor plate and ground as well as the voltage~$V_0$ across that gap we can then calculate the voltage~$V_1$ across the CPW.
    This allows us to determine the energy~$E_{\mathrm{CPW}}$ stored in the CPW fields and finally the contribution~$C_{\mathrm{CPW}}(\tilde s)$ of the CPW to the effective resonator capacitance~$C$.
    See main text for the detailed calculation.
  }
  \label{fig:standing-wave-correction}
\end{figure}

The dimensions of the circuit, in particular the inductive wire length~$s$ of the resonator, are not sufficiently small to neglect the finite microwave wavelength at~\qty{11}{\giga\hertz} when analyzing its resonance properties.
In other words, the wire cannot be treated as a simple lumped element inductance, but -- in combination with the ground planes -- shall be modeled as a distributed element coplanar waveguide of length $\tilde s = s - q$ in series with a short, purely inductive wire of length~$q$, cf.~Fig.~\ref{fig:standing-wave-correction}.
We neglect any capacitance contribution from the wire part with length~$q$, since it is the farthest from the ground planes.

Given the CPW and substrate dimensions (inductive wire width~\qty{20}{\micro\meter}, distance to ground plane~\qty{390}{\micro\meter}, substrate thickness~\qty{330}{\micro\meter}, average dielectric constant of sapphire~$\epsilon_{\mathrm r} = 10$), we find the CPW capacitance per unit length $C' = \qty{56}{\pico\farad\per\meter}$ with an effective permittivity of~$\epsilon_{\mathrm{eff}} = \num{5.06}$~\cite{garg2013microstrip}.
This yields the phase velocity $v_\phi = c_0 / \sqrt{\epsilon_{\mathrm{eff}}} = \qty{1.33e8}{\meter\per\second}$, corresponding to a resonance wavelength $\lambda_0 = 2 \pi v_\phi / \omega_0 = \qty{12.1}{\milli\meter}$ for $\omega_0 = 2 \pi \times \qty{11}{\giga\hertz}$.
Thus, a corresponding quarter-wave standing-wave cavity would have a length of $\lambda_0 / 4 = \qty{3.0}{\milli\meter}$, only about~$\num{2.2} \times \tilde s_{\mathrm{max}}$ with $\tilde s_{\mathrm{max}} = \qty{1.4}{\milli\meter}$ being the largest considered length, confirming that modeling it as a lumped element inductance is not appropriate.

For the characteristic impedance of the CPW~piece with length~$\tilde s$, we find $Z_1 = 1 / v_\phi C' = \qty{135}{\ohm}$ (the kinetic inductance contribution can be neglected for our film thickness~\cite{watanabe1994}; it would only change~$Z_1$ by~\qty{< 1}{\percent}, impacting the coupling strength~$g$ by~\qty{< 0.2}{\percent}).
Seen from the side of the remaining cavity, the short-ended CPW has the input impedance
\begin{equation}
  Z_{\mathrm{in}} = \i Z_1 \tan\Bigl( 2 \pi \frac{\tilde s}{\lambda} \Bigr)
\end{equation}
with $\lambda = 2 \pi v_\phi / \omega$.
We omit the effect of the shunt capacitance~$C_{\mathrm s}$ on the boundary condition or the effective total capacitance as well as any dissipative contributions, since they are small.
The resonance condition can be written as
\begin{equation}\label{eq:res_con}
  \i Z_1 \tan\Bigl( 2 \pi \frac{\tilde s}{\lambda_0} \Bigr) = -\Bigl( \i \omega_0 L_0 + \frac{1}{\i \omega_0 C_0} \Bigr)
\end{equation}
where $C_0$ is the capacitance of the plate only and $L_0$ is the inductance of the circuit without the CPW part~(cf.~Fig.~\ref{fig:standing-wave-correction}).

Now we assume that at the plate capacitance~$C_0$ we have the microwave voltage~$V_0$.
The voltage at the input of the transmission line at that moment~$V_1$ can be obtained from considering everything connected to the capacitance to be a voltage divider with
\begin{align}
  V_1 &= V_0 \frac{Z_1 \tan(2 \pi \tilde s / \lambda_0)}{\omega_0 L_0 + Z_1 \tan(2 \pi \tilde s / \lambda_0)} \\
      &= V_0 \omega_0 C_0 Z_1 \tan\Bigl( 2 \pi \frac{\tilde s}{\lambda_0} \Bigr)
\end{align}
where in the last step we used Eq.~\eqref{eq:res_con}.

A final ingredient is the capacitive energy on the CPW, which can be calculated via
\begin{equation}
  E_{\mathrm{CPW}} = \frac 12 C' V_\lambda^2 \int_0^{\tilde s} \sin^2\Bigl( 2 \pi \frac{x}{\lambda_0} \Bigr) \dif x
\end{equation}
with $V_\lambda$ the hypothetical maximum voltage at $x = \lambda_0 / 4$ and the boundary condition
\begin{equation}
  V_1 = V_\lambda \sin\Bigl( 2 \pi \frac{\tilde s}{\lambda_0} \Bigr) .
\end{equation}
After integration and some algebra we get
\begin{equation}
  E_{\mathrm{CPW}} = V_0^2 (\omega_0 C_0 Z_1)^2 C' \frac{\tilde s - \lambda_0 / 4 \pi \cdot \sin(4 \pi \tilde s / \lambda_0)}{4 \cos^2(2 \pi \tilde s / \lambda_0)}
\end{equation}
from which we conclude the effective capacitance contribution of the CPW to be
\begin{equation}
  C_{\mathrm{CPW}}(\tilde s) = (\omega_0 C_0 Z_1)^2 C' \frac{\tilde s - \lambda_0 / 4 \pi \cdot \sin(4 \pi \tilde s / \lambda_0)}{2 \cos^2(2 \pi \tilde s / \lambda_0)} .
\end{equation}

With \software{Comsol Multiphysics}, we simulated the dc~capacitance~$C_{\mathrm{dc}}$ including the inductive wire, so in order to find the effective total capacitance~$C$ of our cavity, we calculate
\begin{equation}
  C = C_0 + C_{\mathrm{CPW}}(\tilde s) = C_{\mathrm{dc}} - C' \tilde s + C_{\mathrm{CPW}}(\tilde s) ,
\end{equation}
the result of which is plotted in Fig.~\subref{fig:parameter-optimization}{c}.
Note that from our data in Fig.~\subref{fig:parameter-optimization}{c} we can also extrapolate the inductance~$L_0$ as the value of the nearly linear~$L(s)$ for $s = q = \qty{400}{\micro\meter}$, which we find to be $L_0 \approx \qty{0.7}{\nano\henry}$.

\section{Device fabrication}\label{sec:app_devicefab}
\begin{figure}
  \includegraphics{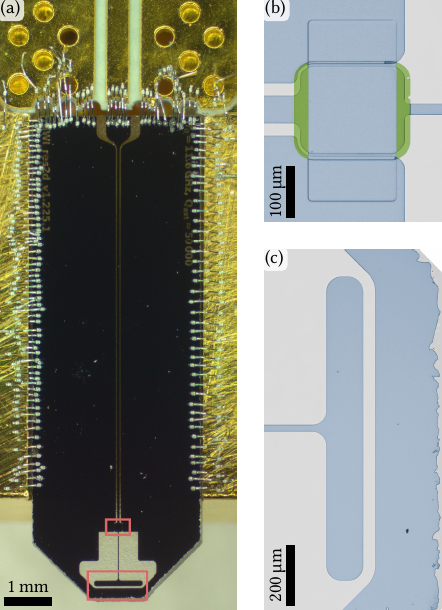}
  \titlecaption{Optical microscopy images of a typical device}{%
    \sublabel{a}~A fully fabricated chip mounted in a sample holder.
    Wire bonds to the PCB at the top as well as to the sample holder body on the sides serve both as electric contacts and as mechanical fasteners holding the chip.
    The niobium appears black due to the lighting conditions, the transparent sapphire shows the color of the sample holder gold plating at the top and that of the microscope table at the bottom.
    Red rectangles mark the position of the shunt and the resonator capacitor shown in \sublabel{b} and~\sublabel{c}, respectively, rotated by~\ang{90}.
    The structures have been colored according to their material:
    Blue for niobium, green for silicon nitride and gray for sapphire.
    Chipping caused by the dicing step at the end of fabrication can be seen at the right edge of~\sublabel{c}.
  }
  \label{fig:optical-images}
\end{figure}

The device fabrication of the trilayer chip is executed in four steps.
These steps are individually described below.
Optical microscopy images of a typical device are shown in Fig.~\ref{fig:optical-images}.

\paragraph*{Step 1: Microwave cavity patterning}
The fabrication starts with dc-magnetron sputtering of \qty{80}{\nano\meter}~thick niobium on top of a two-inch r-cut and single-side-polished sapphire wafer.
The nominal thickness of the wafer is~\qty{330}{\micro\meter}.
For optical lithography, the entire wafer is covered with \mbox{ma-P~1215} photoresist by spin-coating (resist thickness~\qty{\sim 1.5}{\micro\meter}) and loaded into a system for maskless optical lithography ($\lambda_{\mathrm{MLA}} = \qty{365}{\nano\meter}$).
After exposure, the resist is developed in \mbox{ma-D~331/S} for~\qty{35}{\second}, followed by dry-etching of the pattern into the \ce{Nb}~film using reactive ion etching with \ce{SF6}.
For cleaning and removal of the remaining resist, the wafer gets rinsed in multiple subsequent baths of acetone, isopropanol and ultra-pure water.

\paragraph*{Step 2: Dielectric patch for the shunt capacitor}
As a second step, we again perform maskless photo-lithography to define the areas on the wafer which will be covered with \ce{Si3N4} for the parallel-plate shunt capacitor.
After resist development identical to step~1, the wafer with the patterned resist structures is placed inside the vacuum chamber of a plasma-enhanced chemical vapor deposition~(PECVD) system and is covered with \qty{100}{\nano\meter} of \ce{Si3N4}.
Afterwards, an ultrasound-assisted lift-off procedure is performed in acetone which removes the resist and all the \ce{Si3N4} except at the locations of the later shunt capacitors.
Finally, the wafer is rinsed in multiple baths of acetone, isopropanol and ultra-pure water again.

\paragraph*{Step 3: Top plate for the shunt capacitor}
The third step is almost identical to step 2, but instead of PECVD-grown \ce{Si3N4}, a \qty{120}{\nano\meter}~thick layer of niobium is deposited by magnetron sputtering.
To guarantee good galvanic contact between the top and bottom \ce{Nb}~layers on the ground planes, a short dry-etching step with~\ce{SF6} is performed in-situ before the sputtering process.
From our etch rates, we expect a \ce{Nb}~etch depth of~\qty{10}{\nano\meter} and a \ce{Si3N4}~etch depth of~\qty{30}{\nano\meter}, which reduces the final thickness of the dielectric in the shunt capacitor to~\qty{\sim 70}{\nano\meter}.
After liftoff in acetone, the wafer is once more cleaned in multiple baths of acetone, isopropanol and ultra-pure water.

\paragraph*{Step 4: Dicing and mounting}
At the end of the trilayer cavity fabrication, the wafer gets diced into individual, rectangular \qtyproduct[product-units=power]{10 x 3.5}{\milli\meter} chips.
In a second dicing step, the corners at the atom-side of the chip are removed.
Then, one chip at a time is mounted adjacent to a Rogers printed circuit board~(PCB) into a gold-plated copper housing, where it is wire-bonded to microwave feedlines and ground, cf.~Fig.~\subref{fig:optical-images}{a}.
The PCB contains a coplanar waveguide feedline leading to an SMP~connector, where a microwave cable is attached during measurement.
The metal housing consist of gold-plated copper.
After mounting into the measurement setup, the characterization of the device is performed.

\section{Experimental setup}\label{sec:app_setup}

\begin{figure}
  \includegraphics{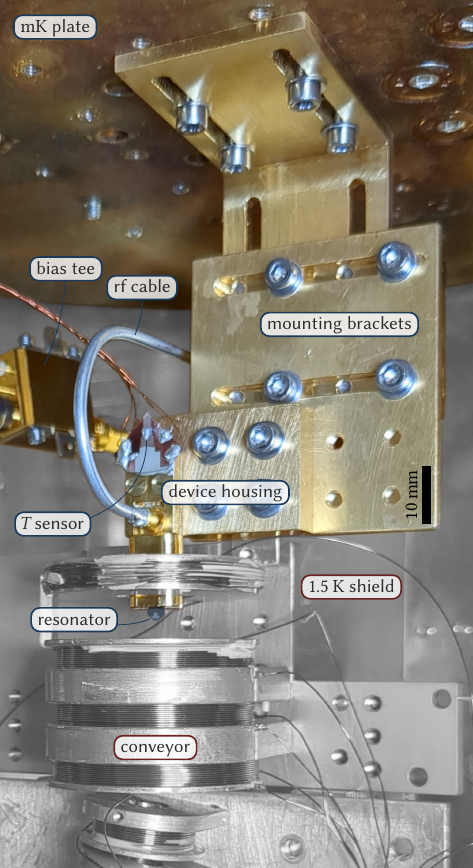}
  \titlecaption{Mounting the device in the dilution refrigerator}{%
    The axis of optical access passes through the center of the quadrupole coils at the top of the conveyor transporting the atoms up from their preparation stage.
    Thus, the resonator must be placed close to that center, such that the atoms can be loaded into the dipole trap and placed in its interaction region.
    We attach the sample holder to the \unit{\milli\kelvin}~plate using a cascade of mounting brackets allowing three-axis movement for precise alignment of the device position.
    The conveyor coils as well as the inner-most heat shield are at~\qty{1.5}{\kelvin} and depicted in gray-scale, all colored parts are at~\unit{\milli\kelvin}.
    A $T$~sensor is placed directly on the copper housing of the device for precise readout of its temperature.
    On the left, the bias~tee can be seen as well as the aluminum rf~cable connecting it to the device~(cf.~Fig.~\subref{fig:T-characterization}{a}).
  }
  \label{fig:mounting-in-fridge}
\end{figure}

The experimental setup in which we have characterized the superconducting cavity and in which we envision to implement the discussed hybrid system is an Oxford Triton dilution refrigerator with optical access and a second \qty{6}{\kelvin}~plate at the bottom for the preparation of ultra-cold atomic ensembles within a magneto-optical trap~\cite{jessen2014}.
Due to the additional cold-atom-specific components in this cryogenic prototype system and the currently large thermal load, the temperature of the plates is somewhat higher than in standard dilution refrigerators.
One consequence is that in our current experiment the lowest temperature of the mixing chamber is $T_{\mathrm{min}} \approx \qty{0.2}{\kelvin}$, and the usual \qty{3}{\kelvin}~plate has a temperature of~\qty{6}{\kelvin}.
In previous cooldowns with much less thermal load, the \unit{\milli\kelvin}~plate in the system reached down to $T_{\mathrm{pr}} < \qty{50}{\milli\kelvin}$ and in the future we will try to reach those temperatures again by optimized thermal engineering and radiation isolation measures.

From the top down to the \unit{\milli\kelvin}~plate, however, the fridge looks nearly identical to standard dilution refrigerators and we have equipped the various temperature stages with the usual microwave components for quantum circuits, cf.~Fig.~\subref{fig:T-characterization}{a}.
The \qtylist[list-units=bracket, list-final-separator={, }]{20; 20; 10}{\deci\bel} attenuator cascade on the input line and the stainless-steel coaxial cables, the circulator and the bias-tee together provide a total input attenuation of roughly~\qty{-70}{\deci\bel} and equilibrate the incoming noise to nearly the base temperature of the system.
Hence, the microwave cavity at~\qty{11.7}{\giga\hertz} is close to its quantum ground state with $n_{\mathrm{th}} \sim \num{0.2}$.
For the rf~connections between the different components on the \unit{\milli\kelvin}~stage we use coaxial cables made of superconducting aluminum.

To mount the chip into the center of the magnetic coils, as shown in Fig.~\subref{fig:T-characterization}{b}, a mounting-bracket cascade made of gold-plated copper allowing three-axis movement is attached to the \unit{\milli\kelvin}~plate, with the microwave box mounted at its end, cf.~Fig.~\ref{fig:mounting-in-fridge}.
This way, the position of the chip can be adjusted for optimal height and alignment with the optical axis as well as avoiding mechanical contact between the device housing at~\unit{\milli\kelvin} and the coils at~\qty{1.5}{\kelvin}.
Note that, in addition to the thermal load mentioned above, thermal radiation emitted by the coils may contribute to the chip not reaching the lowest possible temperature.
To track the chip temperature as precisely as possible, an additional temperature sensor is mounted directly on the copper housing.

\section{Fitting and background-correction of resonance data}\label{sec:app_fitting}
The ideal reflection-response function of a high-$Q$ series RLC~circuit which is coupled to a feedline with characteristic impedance~$Z_0$ by a shunt capacitor~$C_{\mathrm s}$ is given by
\begin{equation}\label{eq:s11_ideal}
  S_{11}^{\mathrm{ideal}} = -1 + \frac{2 \kappa_{\mathrm{ext}}}{\kappa + 2 \i (\omega - \omega_0)}
\end{equation}
with the angular excitation frequency~$\omega$ and the resonance frequency $\omega_0 = 1 / \sqrt{L C_{\mathrm{tot}}}$.
The total capacitance is given by $C_{\mathrm{tot}}^{-1} = C^{-1} + C_{\mathrm s}^{-1}$, $L$ and~$C$ are inductance and capacitance of the uncoupled circuit, respectively.
The internal and external decay rates (linewidths) $\kappa_{\mathrm{int}}$ and~$\kappa_{\mathrm{ext}}$, are given by
\begin{align}
  \kappa_{\mathrm{int}} &= \omega_0^2 R C_{\mathrm{tot}} \\
  \kappa_{\mathrm{ext}} &= \frac{C_{\mathrm{tot}}}{Z_0 C_{\mathrm s}^2}
\end{align}
and the total decay rate by $\kappa = \kappa_{\mathrm{int}} + \kappa_{\mathrm{ext}}$.
The effective resistance~$R$ accounts for all internal losses of the circuit, such as resistive, dielectric or radiative losses.

Due to the cabling and all the microwave components in between the vector network analyzer and the circuit, the ideal reflection is not what we measure, though.
To take frequency-dependent attenuation, the electrical cable length and possible interferences (e.g.~parasitic reflections or imperfect isolation in the circulator) into account, we model the actual measurement signal as
\begin{equation}\label{eq:s11_real}
  S_{11}^{\mathrm{real}} = (a_0 + a_1 \omega + a_2 \omega^2) \e^{\i (\phi_0 + \phi_1 \omega)} \Bigl( 1 - \frac{2 \kappa_{\mathrm{ext}} \e^{\i \theta}}{\kappa + 2 \i (\omega - \omega_0)} \Bigr) .
\end{equation}
Here, we absorbed the minus sign of the ideal response Eq.~\eqref{eq:s11_ideal} into the prefactors and $a_0$, $a_1$, $a_2$, $\phi_0$, $\phi_1$ and $\theta$ are real-valued fit parameters.
During our automated data fitting routine we first remove the absorption resonance from the dataset (leaving a gap in the $S_{11}$-dataset) and fit the remaining $S_{11}$-response in a narrow frequency window (typically five to ten times~$\kappa$) with the background function
\begin{equation}
  S_{11}^{\mathrm{bg}} = (a_0 + a_1 \omega + a_2 \omega^2) \e^{\i (\phi_0 + \phi_1 \omega)} .
\end{equation}
As a result, we obtain preliminary values for $a_0$, $a_1$, $a_2$, $\phi_0$ and $\phi_1$.
Then, we calculate $S_{11}^{\mathrm{real}} / S_{11}^{\mathrm{bg}}$ for the complete dataset and fit the resulting data with
\begin{equation}
  S_{11}^\theta = 1 - \frac{2 \kappa_{\mathrm{ext}} \e^{\i \theta}}{\kappa + 2 \i (\omega - \omega_0)}
\end{equation}
from which we obtain a preliminary set of values for $\omega_0$, $\kappa$, $\kappa_{\mathrm{ext}}$ and $\theta$.
Finally, we use all the preliminary values for $a_0$, $a_1$, $a_2$, $\phi_1$, $\phi_2$, $\omega_0$, $\kappa$, $\kappa_{\mathrm{ext}}$ and $\theta$ as starting parameters to re-fit the original dataset using Eq.~\eqref{eq:s11_real}.
The values thus obtained for $\omega_0$, $\kappa_{\mathrm{int}}$, $\kappa_{\mathrm{ext}}$ or correspondingly for the quality factors $Q_{\mathrm{int}} = \omega_0 / \kappa_{\mathrm{int}}$ and $Q_{\mathrm{ext}} = \omega_0 / \kappa_{\mathrm{ext}}$ are the ones presented in the manuscript, more specifically the ones in Figs.~\ref{fig:T-characterization} and~\ref{fig:Vdc-characterization}.
All the $S_{11}$-datasets in these figures and their corresponding fit curves have been completely background-corrected by dividing the original dataset by $S_{11}^{\mathrm{bg}}$ with the background parameters obtained from the final fit.
Additionally, we have removed the interference angle~$\theta$.

\section{Plate and distance parameters for the 3D flip-chip architecture}\label{sec:app_3D_parameters}
To obtain a series of reasonable capacitor plate geometries for the numerical simulation of the perspective 3D~architecture discussed in Fig.~\ref{fig:3d-design-concept}, we start with what we are keeping constant and fixed.
Firstly, we keep the beam waist in the focus plane at $w_{\mathrm{dp}} = \qty{15}{\micro\meter}$ as well as the laser parameters $\lambda_{\mathrm{dp}} = \qty{800}{\nano\meter}$ and $P_{\mathrm{dp}} = \qty{50}{\milli\watt}$.
As a consequence, the Rayleigh length remains $l_{\mathrm R} = \qty{0.88}{\milli\meter}$.
Secondly, in order to keep the advantage of high field homogeneity in the center of the capacitor plates, we choose the width of the plates to always be equal to their mutual distance~$a = d$.
Furthermore, we want the plate length to always be $l \geq l_{\mathrm{min}} = \qty{250}{\micro\meter}$, such that a cloud with $T_{\ce{Rb}} \sim \qty{1}{\micro\kelvin}$ is comfortably accommodated in a nearly homogeneous field over the complete cloud length.
In order to maintain this field homogeneity also for large $a, d$ we choose $l = 2 a$, given that this is larger than~$l_{\mathrm{min}}$.
Finally, we want to keep the laser power reaching the two chips directly in the Gaussian beam tails to be small.
While giving a precise limit is impossible without careful study of the actual device, we can assume $P_{\mathrm{dir}} \lesssim \qty{0.1}{\nano\watt}$ to be a reasonable limit~\cite{benevides2024, budoyo2016}.
By inverting Eq.~\eqref{eq:P_dir} numerically, we find the ratio~$r_{\mathrm e} = z_0 / w_{\mathrm e} = d / 2 w_{\mathrm e}$ that results in $P_{\mathrm{dir}} = \qty{0.1}{\nano\watt}$ to be $r_{\mathrm e} = \num{3.0}$ which limits us to $d \gtrsim 2 r_{\mathrm e} w_{\mathrm{dp}} = \qty{90}{\micro\meter}$ even with vanishing chip width.
In our simulations, we thus use $d \geq \qty{100}{\micro\meter}$ as a practical limit.

Given these preliminary considerations, we simulate the perspective geometry for plate distances~$d$ ranging from \qty{100}{\micro\meter} to~\qty{600}{\micro\meter} in steps of~\qty{50}{\micro\meter} and calculate all other lengths from this.
For the chip width~$l_{\mathrm{ch}}$ at the position where the laser crosses it, we assume a \ang{45}~tapered chip profile with \qty{100}{\micro\meter} of padding between the capacitor plate and the chip edge, effectively giving~$l_{\mathrm{ch}} = l + a + \qty{250}{\micro\meter}$.
Table~\ref{tab:3D_parameters} summarizes the resulting eleven parameter sets used to obtain the coupling rates shown in Fig.~\subref{fig:3d-design-concept}{c} as well as the maximum possible chip width~$l_{\mathrm{ch}}^{\mathrm{crit}} = l_{\mathrm R} \sqrt{(d / 2 r_{\mathrm e} w_{\mathrm{dp}})^2 - 1}$ that would lead to $P_{\mathrm{dir}} = \qty{0.1}{\nano\watt}$ according to Eq.~\eqref{eq:P_dir} in each case.
These maximum values turn out to be comfortably large for all but the smallest value of~$d$, revealing that further modifications to the design, e.g.~additional grounding around the capacitor plates, are possible without significant spacial restrictions.

\FloatBarrier
\begin{table}[t]
  \titlecaption{Capacitor and chip parameters for the eleven simulation points discussed in Fig.~\ref{fig:3d-design-concept}}{%
    The laser parameters, including the beam waist $w_{\mathrm{dp}} = \qty{15}{\micro\meter}$, the wavelength $\lambda_{\mathrm{dp}} = \qty{800}{\nano\meter}$, the resulting Rayleigh length $l_{\mathrm R} = \qty{0.88}{\milli\meter}$ and the power $P_{\mathrm{dp}} = \qty{50}{\milli\watt}$, are kept constant for all parameter sets.
    \label{tab:3D_parameters}
  }
  \begin{tabular}{ *{2}{S[table-format=3]} S[table-format=4] S[table-format=1.2] S[table-format=2.2] }\toprule
    {$d$ in \unit{\micro\meter}} & {$a$ in \unit{\micro\meter}} & {$l$ in \unit{\micro\meter}} & {$l_{\mathrm{ch}}$ in \unit{\milli\meter}} & {$l_{\mathrm{ch}}^{\mathrm{crit}}$ in \unit{\milli\meter}} \\\midrule
    100 & 100 & 250 & 0.60 & 0.86 \\
    150 & 150 & 300 & 0.70 & 2.36 \\
    200 & 200 & 400 & 0.85 & 3.51 \\
    250 & 250 & 500 & 1.00 & 4.58 \\
    300 & 300 & 600 & 1.15 & 5.62 \\
    350 & 350 & 700 & 1.30 & 6.64 \\
    400 & 400 & 800 & 1.45 & 7.66 \\
    450 & 450 & 900 & 1.60 & 8.66 \\
    500 & 500 & 1000 & 1.75 & 9.66 \\
    550 & 550 & 1100 & 1.90 & 10.66 \\
    600 & 600 & 1200 & 2.05 & 11.65 \\\bottomrule
  \end{tabular}
\end{table}

\bibliography{socathes_resonator}

\begin{thebibliography}{74}
\bibitem [{las()}]{laser-chip-distance-note}%
  \BibitemOpen
  \href@noop {} {}\bibinfo {note} {The distance of~\qty{80}{\micro\meter} seems
  to come out of nowhere here, but it follows from the (setup-restrained) laser
  focus width, the length of the atomic cloud (defined by the dipole trap
  profile and the cloud temperature) and the optimal width of the
  superconducting chip in combination with a maximum value for how many laser
  photons we allow to shine on the chip. Additionally, we take a small safety
  margin of~\qty{\sim 20}{\micro\meter} into account. The details of each of
  these aspects (except for the margin) are discussed in
  Sec.~\ref{sec:dipole-trap} (dipole trap and cloud temperature), in
  Sec.~\ref{sec:parameter-optimization} (capacitor and chip layout/size), and
  in Appendix~\ref{sec:app_laser_on_chip} (calculation of laser light hitting
  the chip).}\BibitemShut {Stop}%
\end{thebibliography}

\end{document}